\pdfoutput=1
\documentclass[usenatbib]{mn2e}
\usepackage{wrapfig}
\usepackage{sidecap}
\usepackage{graphicx}
\usepackage{url}
\usepackage{caption}
\usepackage{longtable}
\usepackage{array}

\title[]{{The SAMI Galaxy Survey: The Low-Redshift Stellar Mass Tully-Fisher Relation}}

\author[Bloom et al.]
{\parbox{\textwidth} 
{J.~V.~Bloom$^{1,2}$,
S.~M.~Croom$^{1,2}$,
J.~J.~Bryant$^{1,2,3}$,
J.R.~Callingham$^{1,2,4}$,
A.~L.~Schaefer$^{1,2,3}$,
L.~Cortese$^{5}$,
A.~M.~Hopkins$^{3}$,
F.~D'Eugenio$^{6}$,
N.~Scott$^{1,2}$,
K.~Glazebrook$^{7}$,
C.~Tonini$^{7}$,
R.~E.~McElroy$^{1,2}$,
H.~Clark$^{1}$,
B.~Catinella$^{5}$,
J.~T.~Allen$^{1}$,
J.~Bland-Hawthorn$^{1}$,
M.~Goodwin$^{3}$,
A.~W.~Green$^{3}$,
I.~S.~Konstantopoulos$^{8}$,
J.~Lawrence$^{3}$,
N.~Lorente$^{3}$,
A.~M.~Medling$^{6,9}$,
M.~S.~Owers$^{2,10}$,
S.~N.~Richards$^{1,2,3}$,
R.~Sharp$^{6}$
}
\vspace{0.4cm} \\
\parbox{\textwidth}{$^{1}$Sydney Institute for Astronomy, School of Physics, University of Sydney, NSW 2006, Australia\\
$^{2}$CAASTRO: ARC Centre of Excellence for All-sky Astrophysics\\
$^{3}$Australian Astronomical Observatory (AAO), 105 Delhi Rd, North Ryde, NSW 2113, Australia\\
$^{4}$CSIRO Astronomy \& Space Science P.O.Box 76, Epping, NSW 1710, Australia\\
$^{5}$International Centre for Radio Astronomy Research, University of Western Australia, 35 Stirling Highway, WA 6009, Australia\\
$^{6}$Research School of Astronomy and Astrophysics, Australian National University, Canberra, ACT 2611, Australia\\
$^{7}$Centre for Astrophysics and Supercomputing, Swinburne University of Technology, PO Box 218, Hawthorn, VIC 3122, Australia\\
$^{8}$Atlassian 341 George St Sydney, NSW 2000\\
$^{9}$Cahill Center for Astronomy and Astrophysics, California Institute of Technology, MS 249-17 Pasadena, CA 91125, USA\\
$^{10}$Department of Physics and Astronomy, Macquarie University, NSW 2109, Australia}
}

\begin{document}

\maketitle

\begin{abstract}

We investigate the Tully-Fisher Relation (TFR) for a morphologically and kinematically diverse sample of galaxies from the SAMI Galaxy Survey using 2 dimensional spatially resolved H$\alpha$ velocity maps and find a well defined relation across the stellar mass range of $8.0<\log(M_{*}/M_{\odot})<11.5$. 

We use an adaptation of kinemetry to parametrise the kinematic H$\alpha$ asymmetry of all galaxies in the sample, and find a correlation between scatter (i.e. residuals off the TFR) and asymmetry. This effect is pronounced at low stellar mass, corresponding to the inverse relationship between stellar mass and kinematic asymmetry found in previous work. For galaxies with $\log(M_*/M_{\odot})<9.5$, $25\pm3\%$ are scattered below the root mean square (RMS) of the TFR, whereas for galaxies with $\log(M_*/M_{\odot})>9.5$ the fraction is $10\pm1\%$

We use `simulated slits' to directly compare our results with those from long slit spectroscopy and find that aligning slits with the photometric, rather than the kinematic, position angle, increases global scatter below the TFR. Further, kinematic asymmetry is correlated with misalignment between the photometric and kinematic position angles. This work demonstrates the value of 2D spatially resolved kinematics for accurate TFR studies; integral field spectroscopy reduces the underestimation of rotation velocity that can occur from slit positioning off the kinematic axis.



\end{abstract}


\section{Introduction}

The Tully-Fisher relation (TFR) \citep{tully1977new} is the fundamental scaling relation between stellar mass \citep{mcgaugh2000baryonic} (originally luminosity) and rotation velocity. It has been shown to hold consistently in the nearby universe for regular disk galaxies [e.g. \citet{bell2001stellar, kassin2006stellar}]. The traditional use of the TFR has been as a distance indicator, via the establishment of the tightest possible scaling relation. This usage requires the exclusion of galaxies outside limited morphology ranges, or that show signs of interaction [e.g. \citet{pierce1992luminosity,bureau1996new,giovanelli1996band,haynes1999band,tully2000distances}].

Scatter below the TFR has been linked to low stellar mass and perturbation using a variety of methods, including optical rotation curves \citep{barton2001tully,kannappan2002physical} and integral field spectroscopy (IFS)\citep{cortese2014sami}. In previous work, we found an inverse relationship between stellar mass and kinematic asymmetry for galaxies in the Sydney--AAO Multi-object Integral field spectrograph (SAMI) Galaxy Survey sample \citep{bloom2016sami}. This result was in agreement with other work, using a variety of metrics for disturbance [e.g. \citet{van1998neutral,cannon2004complex,lelli2014dynamics}], demonstrating that low stellar mass galaxies have complex kinematics, deviating from ordered rotation. Rotation curve measurements of kinematic asymmetry \citep{barton2001tully,garrido2005ghasp} found dwarf galaxies to be both disordered and to have low {measured} rotational velocity. It is also well known that stellar mass is linked to morphological type, with low mass galaxies tending to have irregular morphologies \citep{roberts1994physical,mahajan2015galaxy}.

In the past, spectroscopic measurements were mostly taken using a single fibre or slit \citep{york2000sloan,percival20012Df,driver2009gama}. {There have been extensive TFR studies with single slit and fibre measurements  [\citet{tully1977new}, \citet{courteau1997optical}, \citet{bohm2004tully}, \citet{mocz2012tully} and others]}. Despite the success of fibre and slit-based measurements in determining the TFR, they are vulnerable to potential errors introduced by slit placement {\citep{spekkens2005cusp,oh2011dark,simons2015transition}} and aperture effects. It is also difficult to investigate spatial variation across an extended source using a single slit. Two dimensional spatially resolved kinematics provide a means to circumvent this problem, allowing for greater robustness in kinematic measurements. SAMI is a multiplexed integral field spectrograph, enabling the production of sample sizes of the order of thousands of galaxies on a much shorter timescale than would be possible with a single integral field spectrograph \citep{croom2012sydney}.

Two dimensional spatially resolved spectroscopy has been used to explore the TFR in a variety of ways. The K-band Multi-Object Spectrograph (KMOS) Redshift One Spectroscopic Survey (KROSS) use spatially resolved H$\alpha$ emission to study evolution of the $M_{*}/M_{\odot}$ TFR to $z\sim1$ \citep{tiley2016kmos}. At low redshift, \citet{davis2011atlas3d} use the ATLAS$^{3D}$ sample to show the carbon monoxide (CO) TFR for early-type galaxies. {Recently, the CALIFA Survey have produced a TFR at low redshift using rotation curve fitting to stellar velocity fields \citep{bekeraite2016space}.} The HI TFR has also been thoroughly studied using kinematic maps [e.g. \citet{begum2008baryonic,stark2009first,trachternach2009baryonic,oh2011dark}].

In this work, we show the stellar mass TFR for galaxies in the SAMI Galaxy Survey Sample, and highlight trends with asymmetry and stellar mass. Section~\ref{sec:sami} details the sample selection, instrumentation, data reduction, and methods. Section~\ref{sec:results} presents the TFR for our sample and gives the main relationships between stellar mass, kinematic asymmetry and scatter off the TFR and Section~\ref{sec:scatter} explores the physical and observational causes for scatter below the TFR. We conclude in Section~\ref{sec:conclusion}. We assume a standard cosmology, with $\Omega_m = 0.3,\ \Omega_\lambda = 0.7$ and H$_0 = 70 \mathrm{km/s/Mpc}$.

\section{Sample and Methods}
\label{sec:sami}
\subsection{The SAMI  instrument and SAMI Galaxy Survey sample}

The SAMI instrument uses imaging fibre bundles, or hexabundles \citep{bland2011hexabundles,bryant2014sami}, to take simultaneous integral field spectra for multiple objects. 61 optical fibres comprise each SAMI hexabundle, with each core subtending $\sim1.6$ arcseconds on sky, yielding a total bundle diameter of $15$ arcseconds. {13 hexabundles can be used simultaneously on sky - 12 on galaxies and 1 on a standard star.} The instrument is installed at the prime focus of the Anglo--Australian Telescope (AAT), with fibre cables feeding the double-beamed AAOmega spectrograph \citep{sharp2006performance,croom2012sydney}. 

The 3400 galaxies in the SAMI Galaxy Survey sample were selected from those in the GAMA survey \citep{driver2009gama}, with the addition of 8 galaxy clusters {(Owers et al., in prep)}. The galaxies selected from GAMA sample a broad range in stellar mass and  environment density. The final SAMI Galaxy Survey sample consists of four stellar mass, volume-limited sub-samples, supplemented by additional low mass and filler samples as detailed in \citet{bryant2015sami}.

The SAMI Galaxy Survey data reduction pipeline produces datacubes for each galaxy with adequate signal to noise. For a full description of the SAMI Galaxy Survey data reduction pipeline, we refer the reader to \citet{sharp2015sami}. {For the early SAMI data release, see \citet{allen2015sami}, and for the upcoming full first data release, see Green et al. (in prep.).}

This work builds on that in \citet{bloom2016sami}, and uses similar data products from the SAMI Galaxy Survey including spectral fits using {\small LZIFU}{\sc}. {\small LZIFU}{\sc} is a spectral fitting pipeline written in Interactive Data Language ({\small IDL}{\sc}) that performs flexible emission line fitting of IFS data cubes. The  {\small LZIFU}{\sc} pipeline produces 2D emission line strength and kinematic maps for user-assigned lines. For a more detailed explanation of the {\small LZIFU}{\sc} pipeline, see \citet{ho2016sami}.

At the start of this work, 827 galaxies had been observed and processed through the  {\small LZIFU}{\sc} pipeline. In previous work, we used a H$\alpha$ signal to noise (S/N) cut of 10 to exclude noisy spaxels from the velocity fields output by  {\small LZIFU}{\sc}. Here, we relax the S/N cut to 6, in order to increase the size of the sample. The median velocity error did not increase, {and the decrease in mean spaxel S/N, from $56.4\pm0.1$ to $48.6\pm0.1$ was deemed to be acceptably small}. It was thus found that this increased the number of galaxies in the sample without compromising data quality. 813 galaxies met the S/N cut requirements and yielded results from kinemetry, a 230$\%$ increase on the sample size in  \citet{bloom2016sami}. {As in \citet{bloom2016sami} the H$\alpha$ S/N cut resulted in a sample with a broad range of morphological types, but a bias towards late-type galaxies. }

\subsection{TFR and asymmetry measurements}
\subsubsection{Stellar mass}
\label{sec:vrot_sm}
The stellar mass measurements are from the GAMA Survey catalogue StellarMasses \citep{taylor2011galaxy}, and are based on $ugriz$ SEDs constructed from matched aperture photometry. Typical errors {on individual stellar mass values} are 0.1dex. {This is an internal error to our data set and does not account for systematic offsets in stellar mass measures between our data and other papers that adopt different methods of stellar mass estimation, potentially using different model stellar populations.}

\subsubsection{Inclination}
\label{sec:inc}
In order to correct for velocity projection, we calculate galaxy inclination such that:
\begin{equation}
cos^{2}(i)=\frac{(1-\eta)^{2}-(1-\eta_{max})^2}{1-(1-\eta_{max})^2} ,
\end{equation}
where $i$ is the angle of inclination, $\eta$ is the ellipticity and $\eta_{max}=0.8$ is taken as the ellipticity exhibited by an edge-on disk. Ellipticities are calculated using 400 arcsec r-band cutout image from the Sloan Digital Sky Survey (SDSS). The input image is processed with SExtractor \citep{bertin1996sextractor} to mask out neighbouring objects. PSFEx \citep{bertin2011automated} is used to build a model point spread function (PSF) at the location of the galaxy centre. The ellipticity is then found as a light-weighted mean for the whole galaxy, using the Multi-Gaussian Expansion (MGE) technique \citep{emsellem1994multi} and code from \citet{cappellari2002efficient}. For more detail, we refer to D'Eugenio et al. (in prep.).

\begin{figure}
\centering
\includegraphics[width=9cm]{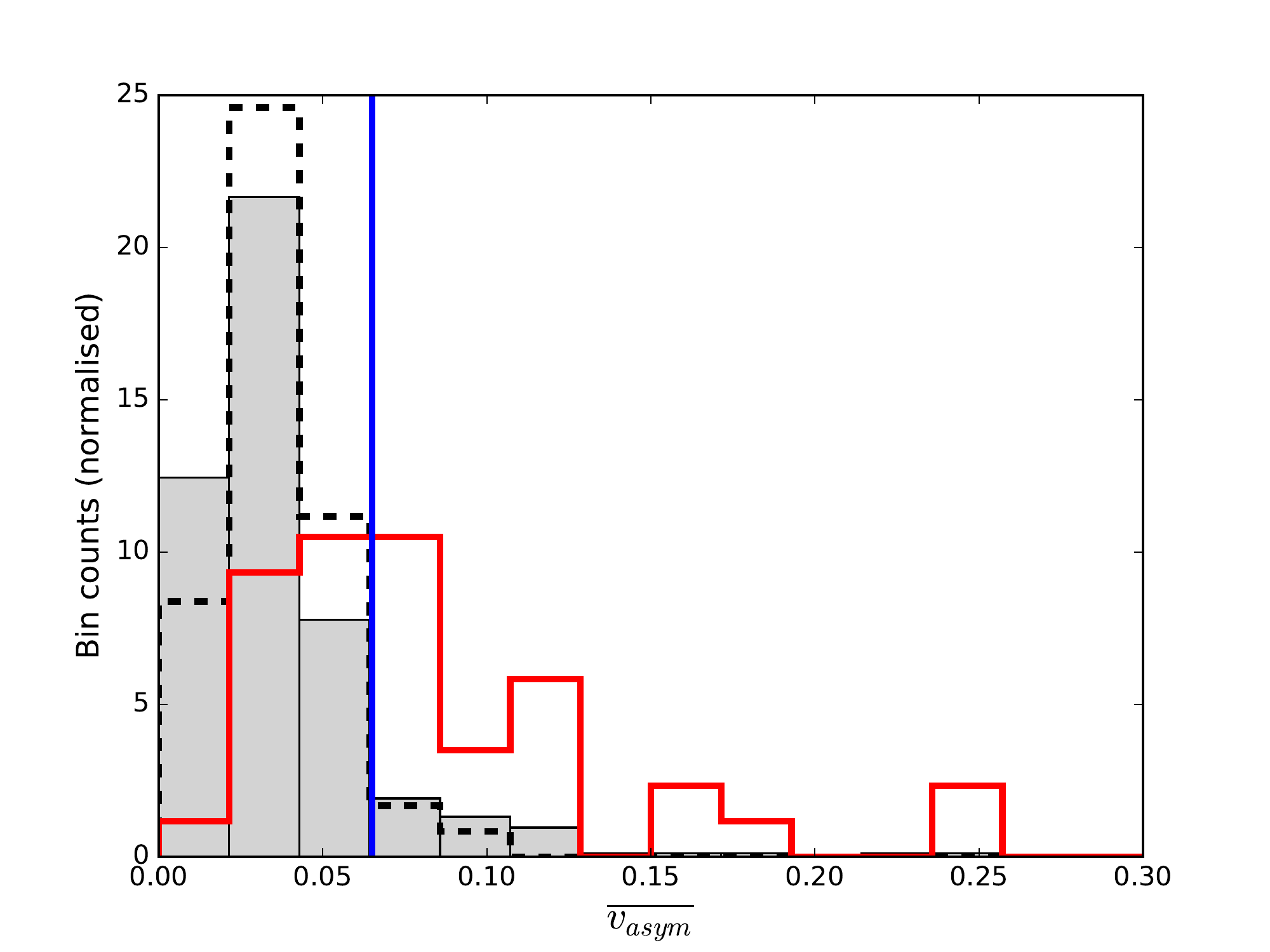}
\caption{Histogram of $\overline{v_{asym}}$ for the full sample in this work (grey), galaxies visually classified as normal (black, dashed) and perturbed (red) in previous work. {The asymmetry cutoff from  \citet{bloom2016sami} is shown in blue, for reference.} Visually classified asymmetric and normal galaxies are seen to have distinct distributions of $\overline{v_{asym}}$.}
\label{fig:asym_hist}
\end{figure}

\begin{figure}
\centering
\includegraphics[width=9cm]{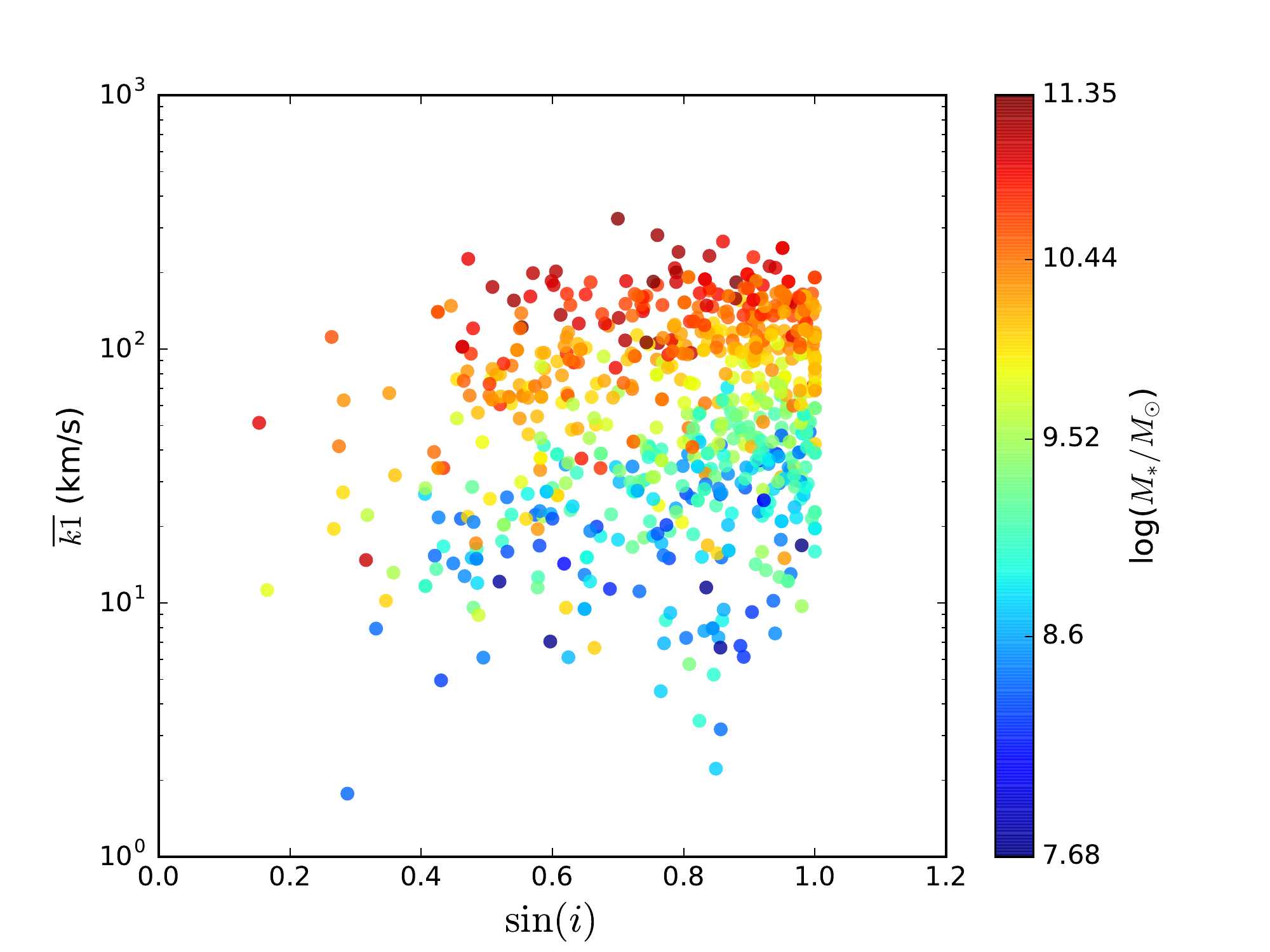}
\includegraphics[width=9cm]{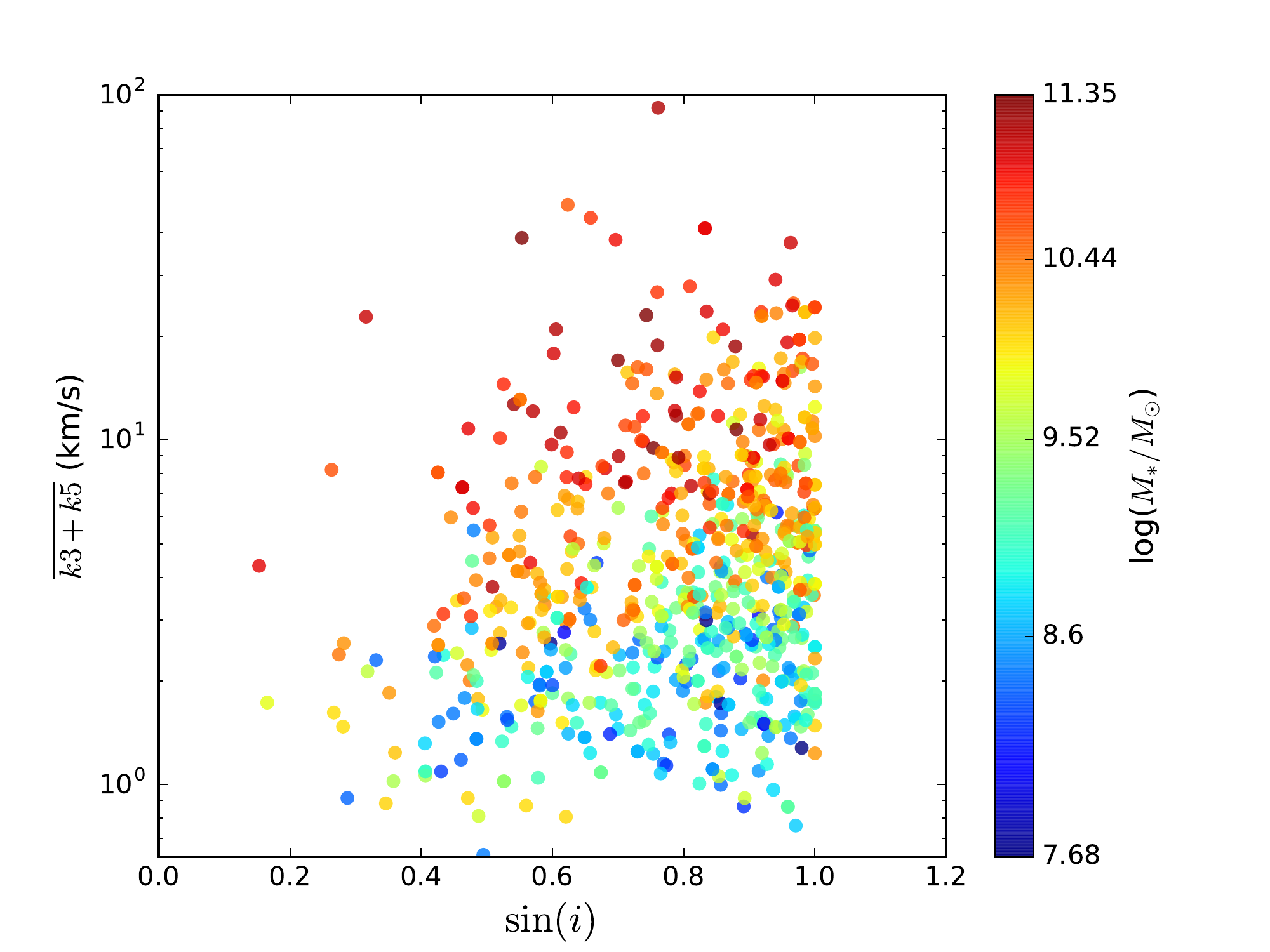}
\caption{Inclination correction ($\sin(i)$) versus $\overline{k1}$ and $\overline{k3+k5}$, with points coloured by $\log(M_*/M_{\odot})$. As expected, $\sin(i)$ is correlated with $\overline{k1}$ and $\overline{k3+k5}$. {Note that $\overline{k1}$ represents bulk rotation, whereas $\overline{k3+k5}$ carries kinematic asymmetry.}}
\label{fig:i_k}
\end{figure}

\subsubsection{Kinemetry}
\label{sec:kinemetry}

Kinemetry is an extension of photometry to the higher order moments of the line of sight velocity distribution (LOSVD)\footnote{The kinemetry code is written in IDL, and can be found at \url{http://davor.krajnovic.org/idl/} \citep{krajnovic2006kinemetry}.}. It was developed as a means to quantify asymmetries in stellar velocity (and velocity dispersion) maps. Such anomalies may be caused by internal disturbances or by external factors, namely interactions \citep{krajnovic2006kinemetry}.

In  \citet{bloom2016sami}, we defined a quantitative asymmetry measure, $\overline{v_{asym}}$, derived from kinemetry \citep{krajnovic2006kinemetry} and justified by a thorough by-eye classification based on SDSS imaging. Fig.~\ref{fig:asym_hist} shows  $\overline{v_{asym}}$ for the full sample in this work (grey), galaxies visually classified as normal (black, dashed) and perturbed (red) in \citet{bloom2016sami}. Visually asymmetric and normal galaxies are seen to have distinct distributions of $\overline{v_{asym}}$.

The kinemetry algorithm fits a series of regularly spaced, concentric ellipses, from which radial moment values are taken. The ellipses are fit with parameters defined by the galaxy centre, kinematic position angle (PA) and photometric ellipticity. We note a change from \citet{bloom2016sami}, in which we used the photometric PA. The ellipticity is taken from MGE fitting, as described in Section~\ref{sec:inc}. The kinematic PA is calculated by the FIT\_KINEMATIC\_PA routine from \citet{krajnovic2006kinemetry}. {We use the photometric centre to define the centre of the fitted ellipses. We fit a 2-dimensional Gaussian to the  SAMI Galaxy Survey $r$-band continuum flux maps and took the centroid of the location of the 25 brightest spaxels in a 6x6 pixel area around the centre of the fitted Gaussian. We did not use the the H$\alpha$ emission maps because they contain clumps of star formation and other features, which make determining the centre from these maps potentially unreliable.} 

The combination of the fitted ellipses is used to construct individual moment maps. As defined in  \citet{bloom2016sami}, following \citet{shapiro2008kinemetry}, the quantitative measure of kinematic asymmetry is found as the median asymmetry over the radii of the fitted ellipses from kinemetry:
\begin{equation}
\overline{v_{asym}}=\overline{\left({\frac{k3+k5}{2k1}}\right)}
\end{equation}
where $k1, k3, k5$ are the coefficients of the Fourier decomposition of the first, third and fifth order moments of the line of sight velocity distribution, respectively. Regular rotation is carried in the $k1$ term, and asymmetry in the higher order terms. {That is, a perfect disk with entirely regular rotation would have no power in the higher order terms, but would be completely represented by $k1$. We do not include $k0$, $k2$ or $k4$ as these terms are used for studying the velocity dispersion map. For a full explanation of our use of kinemetry, see  \citet{bloom2016sami}.}

Fig.~\ref{fig:i_k} shows inclination correction against $\overline{k{1}}$ and $\overline{k{3}+k{5}}$ with points coloured by stellar mass. As expected, $\overline{k1}$ and inclination correction are correlated, with a Spearman rank correlation test giving $\rho=0.2$ and a probability {of no correlation} $p=9.2\times10^{-8}$. Similarly, $\overline{k3+k5}$ and inclination correction have correlation coefficients $\rho=0.2, p=3.4\times10^{-8}$. If perturbations are randomly oriented with respect to the disk, for example due to turbulence, we would not expect a correlation with inclination. The observed correlation implies that perturbations tend to be in the plane of the disk. Given that both $\overline{k1}$ and $\overline{k3+k5}$ are correlated with $\sin(i)$, we do not apply an inclination correction to $\overline{v_{asym}}$.


Due to the re-sampling process used to make the datacubes, the noise in neighbouring spaxels is correlated.  This effect is negligible for spaxels spaced more than $\sim$2.5 spaxels apart. A full explanation of how covariance is handled in SAMI Galaxy Survey data can be found in \citet{sharp2015sami}. In \citet{bloom2016sami} we forced the separation between kinemetry ellipses to be $>2.5$ spaxels. This avoided error underestimation, but reduced the number of ellipses such that it was not possible to extract rotation curves with sufficient detail for our purposes here. Accordingly, we allow the spacing between ellipses to be as low as 1 spaxel and account for covariance by bootstrapping errors. In most cases, perturbation in the velocity field, rather than error on $k1$ terms, dominates the deviation of the fits from perfect, regular rotation.

\subsubsection{$V_{rot}$}
\label{sec:vrot}
The $V_{rot}$ measurements are obtained from the radial $k1$ values fit by the kinemetry package \citep{krajnovic2006kinemetry},  with the $k1$ term carrying the bulk rotation of the disk. Using $k1$ also has the advantage of removing most disturbance, as it traces only the first order moment of the velocity map.

We applied the Bayesian model inference routine outlined in \citet{Callingham2015} to fit the rotation curves of the various galaxies in this study as arctan curves, using the model from \citet{courteau1997optical}:
\begin{equation}
V=V_{{circ}}\times\bigg(\frac{2}{\pi}\bigg)\times \mathrm{arctan}\bigg(\frac{x}{r_t}\bigg) + c
\label{equation:arctan}
\end{equation}
where $V_{circ}$ is the asymptotic velocity, $x$ the distance from centre along the major axis, $r_t$ the transition radius between the rising and flat part of the rotation curve and $c$ is an offset term occurring when the fits do not pass through $(0,0)$. Note that $c<5\mathrm{km/s}$ in all fits, and is typically $\sim1\mathrm{km/s}$. We take the rotation velocity, $V_{rot}$ as $V$ at $2.2r_{e}$, where $r_e$ is the effective radius from MGE fitting, because this is the location of peak rotational velocity of a pure exponential disk \citep{courteau1997optical}. {Choosing $2.2r_{e}$ as our cutoff value also ensured that we did not trace the fitted curve as far beyond the spatial coverage of the data as would be the case using the asymptotic velocity.  In some cases, we do not reach out to $2.2r_{e}$, but we have found that these galaxies do not have a different distribution of $V_{rot}$ values from the rest of the sample and their $k1$ rotation curves are still well fit by Equation~\ref{equation:arctan}.}

The posterior probability density functions of the model parameters of Equation~\ref{equation:arctan} were sampled using the Markov chain Monte Carlo (MCMC) algorithm $\tt{emcee}$ \citep{ForemanMackey2013}. As an affine-invariant ensemble sampler, $\tt{emcee}$ is suited to fitting the rotation curves because it is not significantly impacted by covariance between the parameters \citep{Goodman2010}. A Gaussian likelihood function was maximised using $\tt{emcee}$, with physically motivated uniform priors (i.e. no negative distances, $r_t>0$), by applying the simplex algorithm to direct the walkers \citep{Nelder1965}. The data were assumed to be Gaussian and independent in this process. The uncertainties derived from the method represent the difference between the mean of the posterior distribution and the 16$^{\mathrm{th}}$ and 84$^{\mathrm{th}}$ percentiles. 

\begin{figure*}
\centering
\includegraphics[width=15cm]{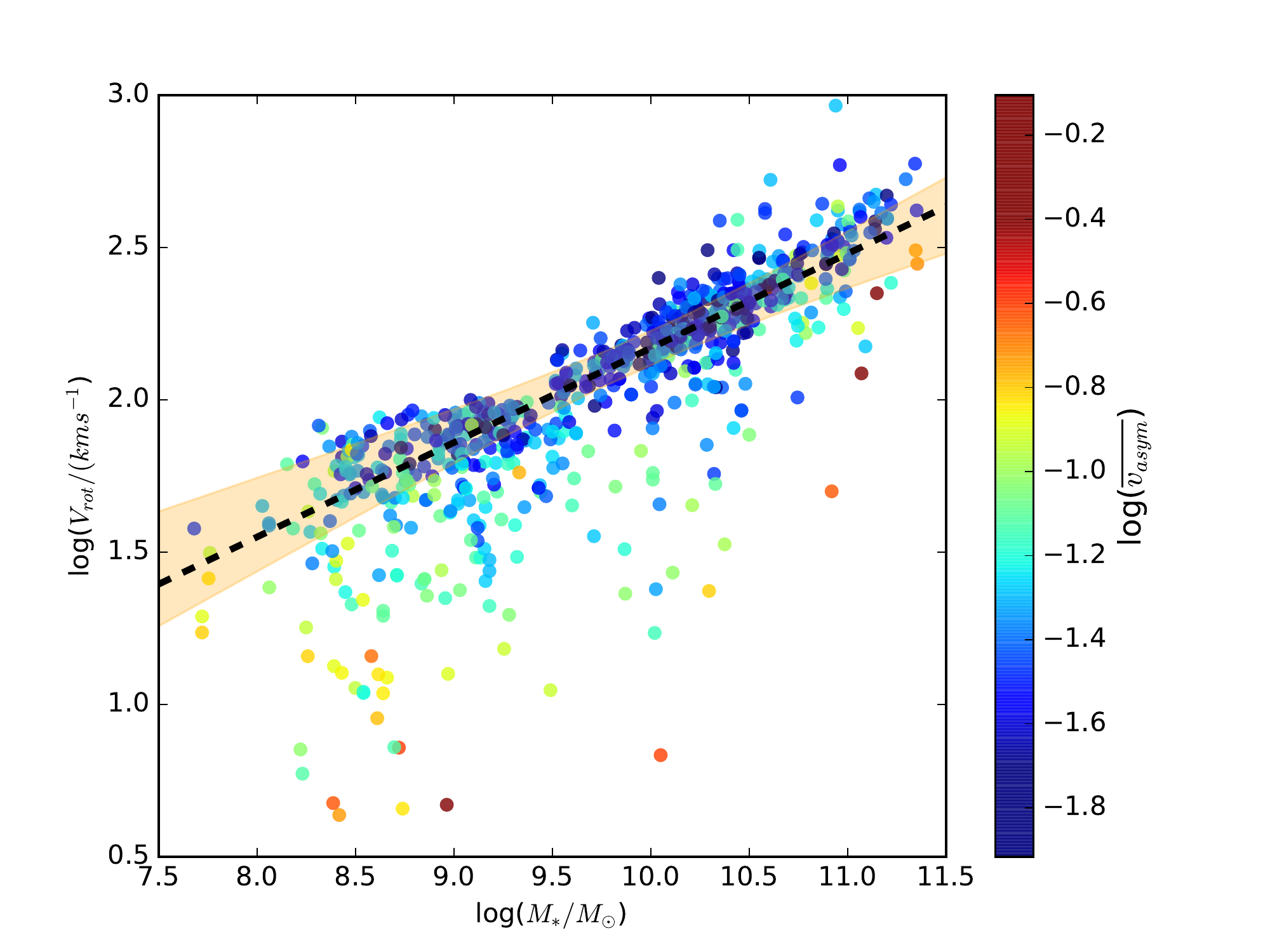}
\caption{The TFR, showing log($V_{rot}/km s^{-1}$) at 2.2$r_e$ against log($M_{*}/M_{\odot}$), with points coloured by $\log(\overline{v_{asym}})$. $V_{rot}$ is calculated from the radial $k1$ moment values produced by kinemetry. The TFR extends across the stellar mass range of the sample, although scatter increases at low stellar mass. Scatter is also correlated with asymmetry. The fit to low-asymmetry galaxies in Equation~\ref{equation:line} is shown in black (dashed), with 1$\sigma$ errors on the fit indicated by the orange region.} 
\label{fig:twod_tfr_asyms}
\end{figure*}

\section{The SAMI Galaxy Survey TFR}
\label{sec:results}

\subsection{2D Spatially resolved TFR}
Using the rotation curves generated by kinemetry (i.e. the $k1$ radial values), we produce a `best case' TFR in Fig.~\ref{fig:twod_tfr_asyms}, with stellar masses from the GAMA Survey StellarMasses catalogue \citep{taylor2011galaxy} and points coloured by kinematic asymmetry. Rotation velocities were fit as in Section~\ref{sec:vrot}, with 729 of 813 galaxies fit successfully and passing quality control. Galaxies were rejected by the fitting routine if there were fewer than 5 points in the $k1$ radial array with errors $<\mathrm{12km/s}$. We choose $\mathrm{12km/s}$ because beyond this value the scatter in the fitted velocities increases substantially. The median number of input points was 13. After error cuts, the median number of input $k1$ values was 12. Fits failed quality control if there was no turnover point in the data, {and the data did not extend out to $2.2r_e$} (leading to unphysical fitted asymptotic velocities $>1000\mathrm{km/s}$) or if $\chi^2>200$. We discuss quality control and fitting further in Section~\ref{sec:simslits}. Henceforth, galaxies with fits failing quality control or with too few included spaxels are considered `not fit'.  

The median $\overline{v_{asym}}$ of the not fit and the fit sample are not statistically offset at $0.050\pm0.01$ and $0.049\pm0.002$, respectively. The highest value in the not fit sample is $\overline{v_{asym}}=0.87$, whereas in the fit sample it is $\overline{v_{asym}}=0.78$. 


The other influence on whether galaxies could be fit for circular velocity is H$\alpha$ S/N. {However, the median H$\alpha$ S/N for those that could have their velocity fit, and those that could not be fit are within a standard error of each other: $26.0\pm1.2$ and $25.8\pm2.6$, respectively. Similarly, the median number of spaxels is also within uncertainties:  $590.0\pm6.2$ and $617.2\pm12.9$.} 
Table~\ref{table:comb} details the fractions of galaxies successfully fit and gives the numbers of galaxies that were excluded from the sample for various reasons.

Poor spatial coverage and low H$\alpha$ S/N are shown to cause galaxies to be rejected by our fitting routine for circular velocity. However, given that the number of rejected galaxies is $\sim10\%$, and that the median $\overline{v_{asym}}$, number of spaxels and median S/N remain unchanged, we conclude that the influence of these factors is small and does not cause significant bias in our sample, despite a weak tendency to reject the most asymmetric galaxies.



The TFR in Fig.~\ref{fig:twod_tfr_asyms} extends through the low stellar mass region of the sample. We stress that, unlike TFRs produced as distance indicators, we include all galaxies that satisfy our fitting and quality control procedures.

To define a fiducial TFR , we exclude asymmetric galaxies (being conservative, we set the cutoff to $\overline{v_{asym}}<0.04$) from the sample used to fit the TFR line because, in general, their gas velocities are not expected to follow the same relation as normal galaxies and they are seen to be outliers. We note that, within the range up to $\overline{v_{asym}}<0.06$, there is no significant change in slope of or scatter around the TFR. Using only this low-asymmetry sample of galaxies, we fit the TFR for our sample:
\begin{equation}
\mathrm{log}(V_{rot}/kms^{-1})=0.31 \pm {0.0092} \times \mathrm{log}(M_{*}/M_{\odot})- 0.93 \pm 0.10
\label{equation:line}
\end{equation}
with an RMS of 0.15 dex in $\log(V_{rot}/kms^{-1})$ for the whole sample. {The RMS for galaxies with $\overline{v_{asym}}<0.04$ is 0.091dex}. Due to the covariance between SAMI Galaxy Survey spaxels, we anticipate that the errors on individual $V_{rot}$ values will be underestimated by the fitting code. Accordingly, in order to obtain errors on fitted TFR parameters, we bootstrap the calculation. The bootstrapping involved 1000 iterations of adding Gaussian noise, defined by scatter from the best fit, to the sample and recalculating the linear fit parameters, with the final error taken from the scatter in the fitted parameters over all the iterations. Henceforth we define scatter as the difference between the expected velocity from the power law fit and the observed velocity, for a given stellar mass.

As has been seen in previous work [e.g. \citet{barton2001tully,kannappan2002physical}], scatter below the TFR increases at low stellar masses ($\log(M_*/M_{\odot})<9.5$). We find, using a Spearman rank correlation test of scatter to stellar mass, $\rho=-0.21, p=2\times10^{-9}$.

\begin{figure}
\centering
\includegraphics[width=9cm]{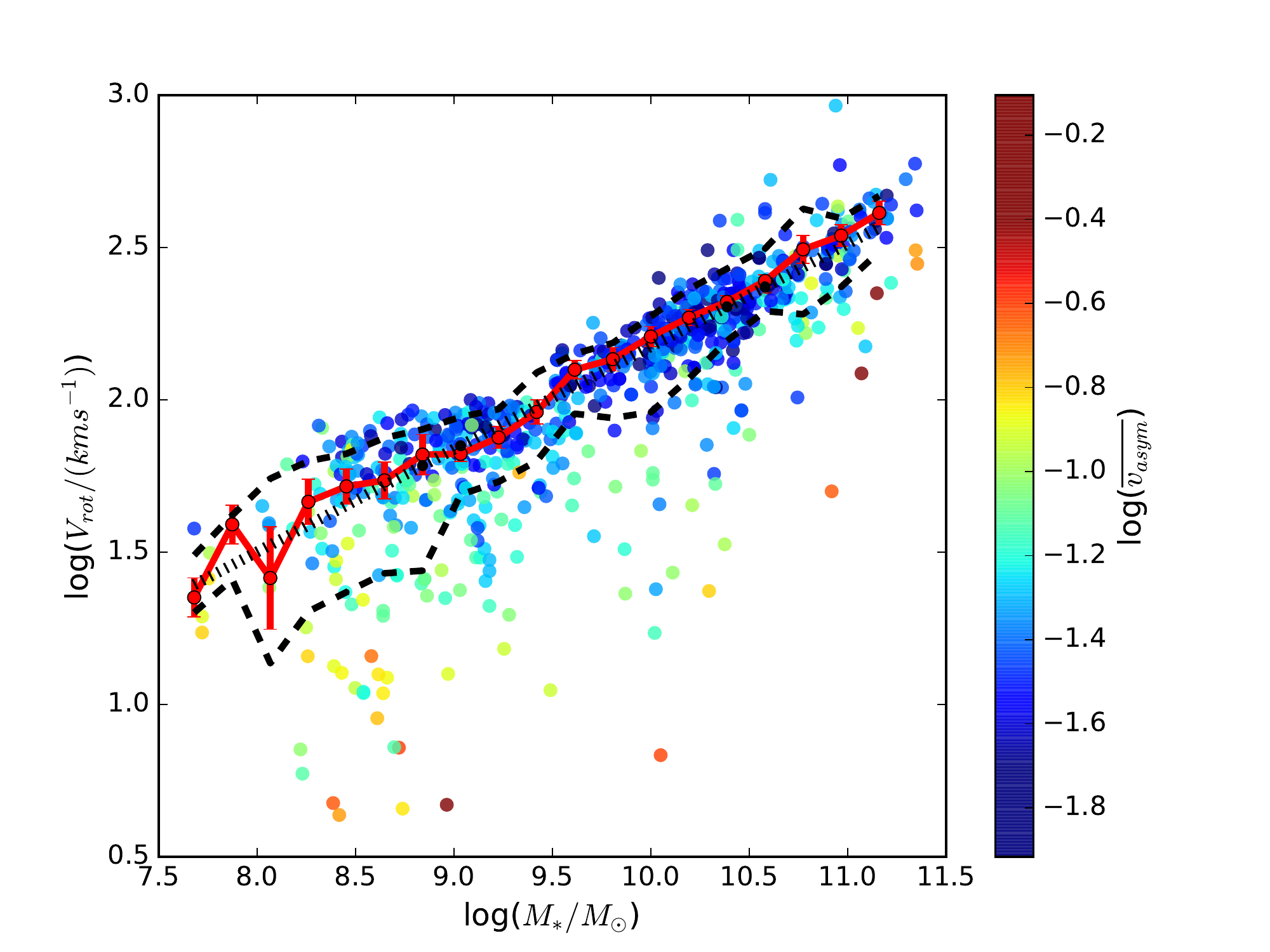}
\caption{The TFR as in Fig.~\ref{fig:twod_tfr_asyms}, with median rotation velocity in stellar mass bins (red line) and RMS in stellar mass bins for all galaxies (dashed, black). The TFR line from Equation~\ref{equation:line} is shown in dotted black.}
\label{fig:tfr_scatter}
\end{figure}

\begin{figure*}
\includegraphics[height=9cm,width=\textwidth]{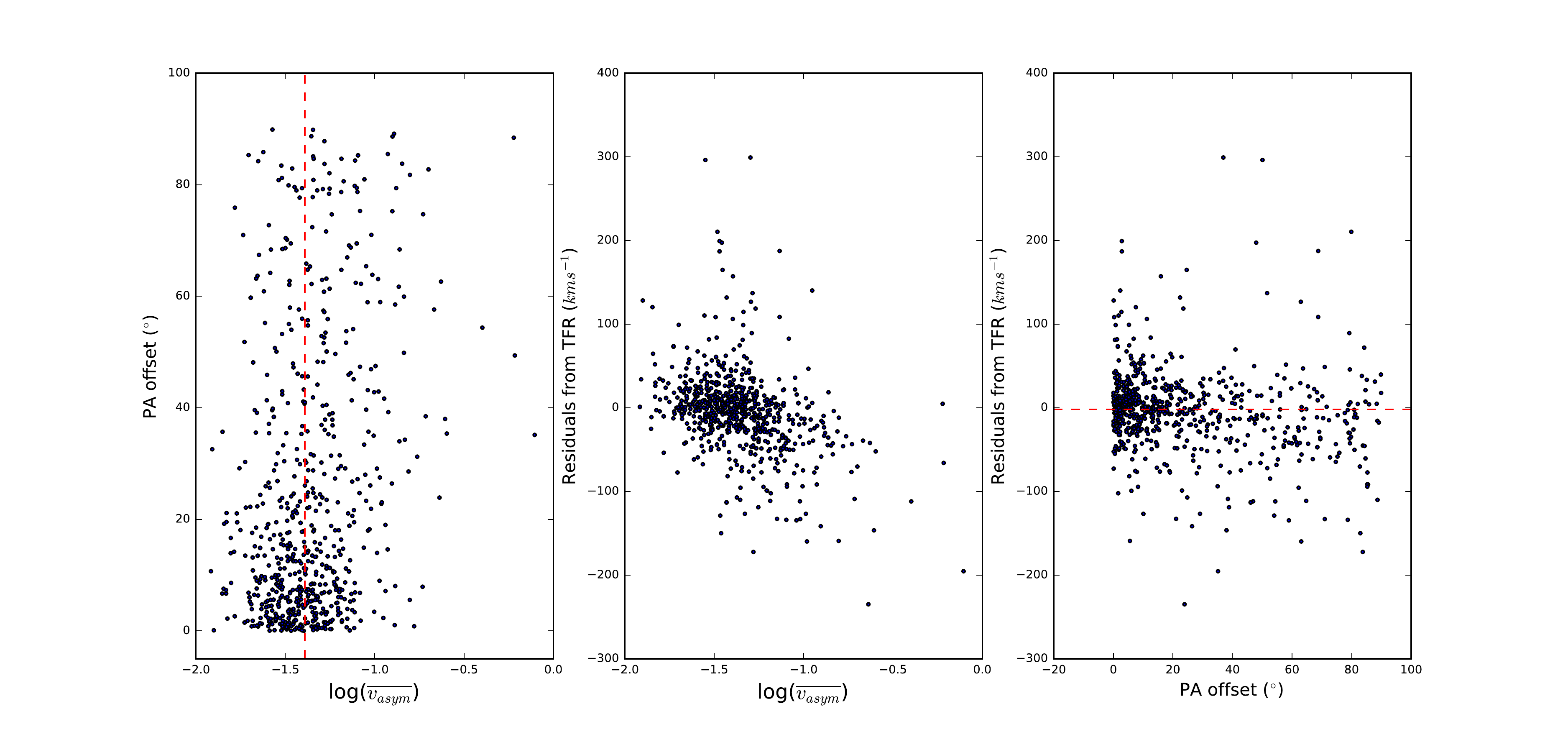}
\caption{From left: $\overline{v_{asym}}$ against PA offset, $\overline{v_{asym}}$ against scatter off the 2D spatially resolved TFR, and PA offset against scatter off the spatially resolved TFR. The correlation between PA offset and $\overline{v_{asym}}$, and the inverse relationships between $\overline{v_{asym}}$ and scatter off the TFR and PA offset and scatter off the TFR can be seen. However, the strongest trend is clearly the inverse correlation between scatter off the TFR and $\overline{v_{asym}}$. Red dashed lines are included at the median of $\overline{v_{asym}}$ in the leftmost plot and TFR residual in the rightmost plot as visual aids to show the correlations.}
\label{fig:p_a_s}
\end{figure*}

Scatter below the TFR is correlated with kinematic asymmetry, with $\rho=0.44, p=4\times10^{-38}$, as can be seen in Fig.~\ref{fig:twod_tfr_asyms} (and explicitly in Fig.~\ref{fig:p_a_s}). Fig.~\ref{fig:tfr_scatter} more explicitly shows the scale of the scatter in our TFR. Of the 153 galaxies scattered below the root mean square (RMS) line, calculated for the whole asymmetry range, in Fig.~\ref{fig:tfr_scatter} (lower dashed line), $49\%$ have $\overline{v_{asym}}>0.065$. This is a much higher proportion of asymmetric galaxies than is found for galaxies above the RMS line ($11\%$). The median $\log(V_{rot}/kms^{-1})$ in bins of stellar mass is also shown in red in Fig.~\ref{fig:tfr_scatter}. It is consistent with the fitted line over the full stellar mass range of the sample. There is more negative scatter in the low than higher mass cases: $25\pm3\%$ of galaxies with $\log(M_*/M_{\odot})<9.5$ lie below the RMS line, compared to $10\%\pm1$ of galaxies with $\log(M_*/M_{\odot})>9.5$. This result is in agreement with \citet{krajnovic2011atlas3d}, \citet{oh2016sami} and others.

An offset between the photometric and kinematic PA is a known indicator of interaction. As in \citet{fogarty2015sami}, this offset is calculated as:
\begin{equation}
\sin\Psi=|\sin(\rm{PA_{phot}-PA_{kin}})|
\end{equation}
where $\Psi$ is the offset, and $\rm{PA_{phot}}$ and $\rm{PA_{kin}}$ are the photometric and kinematic PAs, respectively. The offset lies between $0^{\circ}-90^{\circ}$, and is insensitive to $180^\circ$ differences in the PAs. The kinematic PA found as in Section~\ref{sec:kinemetry} and photometric PA from single MGE fits to the SDSS $r$-band images in the GAMA Survey DR2 catalogue SersicCat \citep{kelvin2012galaxy}.  

All galaxies scattered below the RMS line have an offset between the kinematic and photometric PA $>15^{\circ}$.

\begin{figure*}
\centering
\includegraphics[width=18cm]{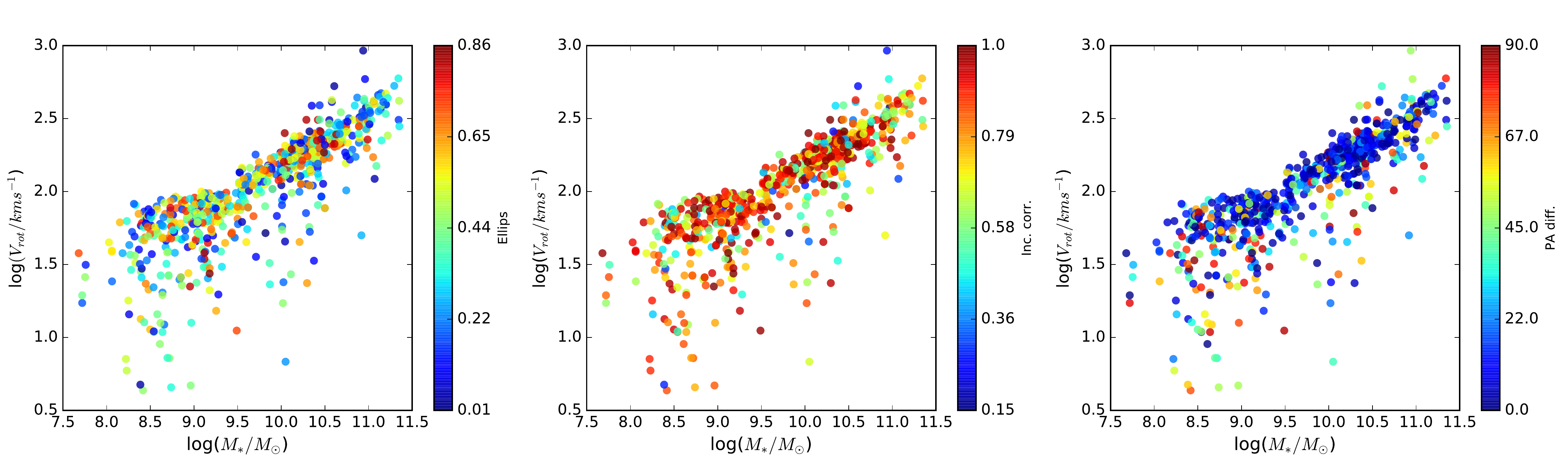}
\caption{The TFR as in Fig.~\ref{fig:twod_tfr_asyms}, with points coloured (left to right) by ellipticity, inclination correction and offset between the kinematic and photometric PA. There is no trend in the first two cases, but a there is one in the third. This indicates that scatter off the TFR is not associated with geometric correction, but it is linked to a physical process: the offset between PAs. }
\label{fig:bias}
\end{figure*}

Fig.~\ref{fig:bias} shows the TFR as in Fig.~\ref{fig:twod_tfr_asyms}, but with points coloured by inclination correction, ellipticity and offset between the kinematic and photometric PAs. There is no relationship between inclination correction and scatter off the TFR either upwards or downwards ($\rho=-0.022, p=0.56$), or between ellipticity and scatter off the TFR ($\rho=-0.021, p=0.55$). There is, however, a correlation between PA offset and scatter, with $\rho=0.18, p=6.3\times10^{-7}$. We discuss this relationship further in Section~\ref{sec:offpa}.

{We note that the population of galaxies with ellipticity$<0.1$ have slightly higher dispersion around the TFR line than the rest of the sample, with an RMS of 0.21 in $\log(V_{rot}/kms^{-1})$, compared to 0.15 for the whole sample. However, these galaxies do not exclusively lie off the TFR, and represent only $4\%$ of the sample, so do not contribute significantly to statistical scatter. Given that we aim to be as complete as possible in our analysis, we do not perform an ellipticity cut.}

\begin{figure*}
\centering
\includegraphics[width=7cm]{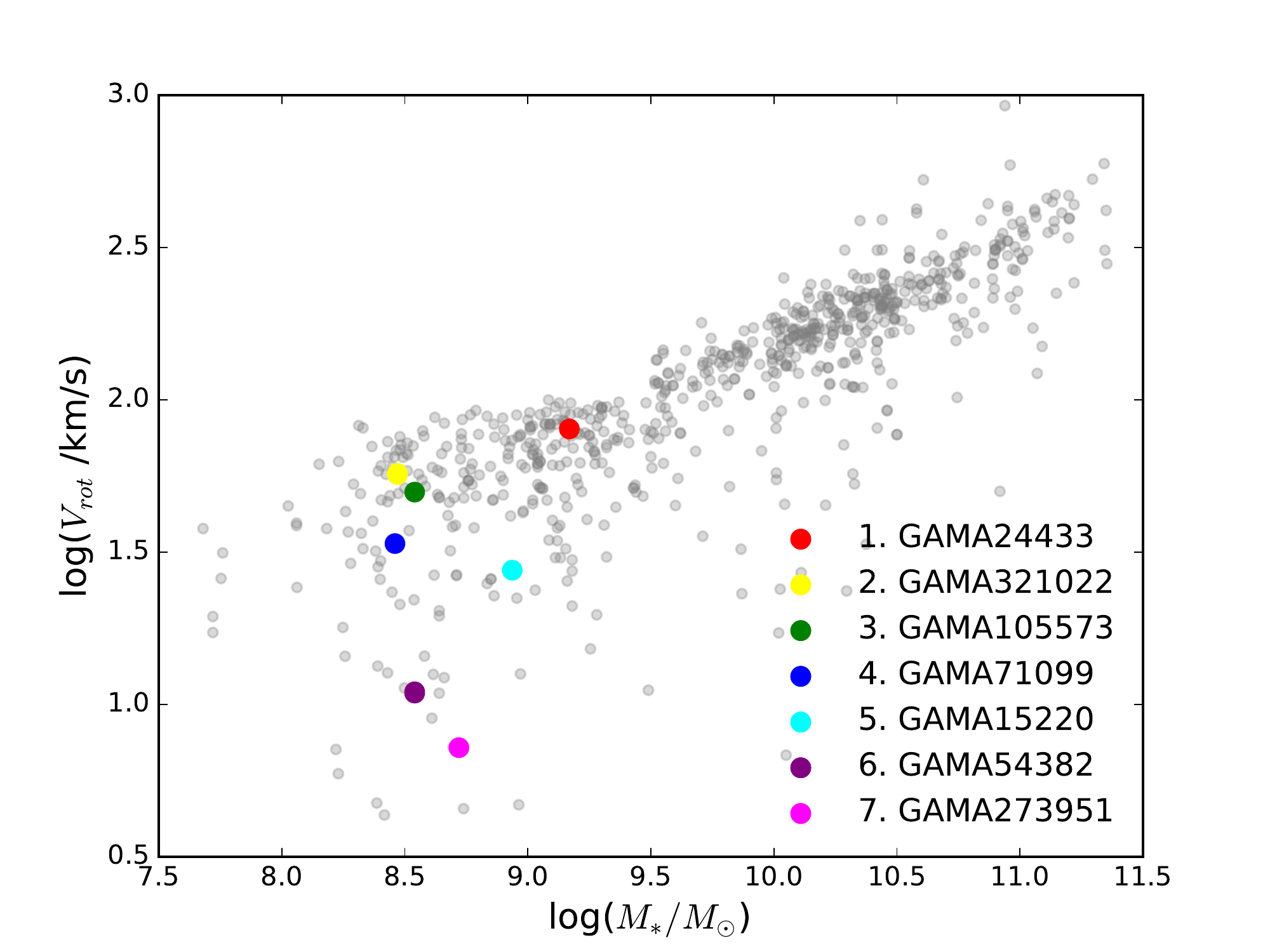}
\includegraphics[width=7cm]{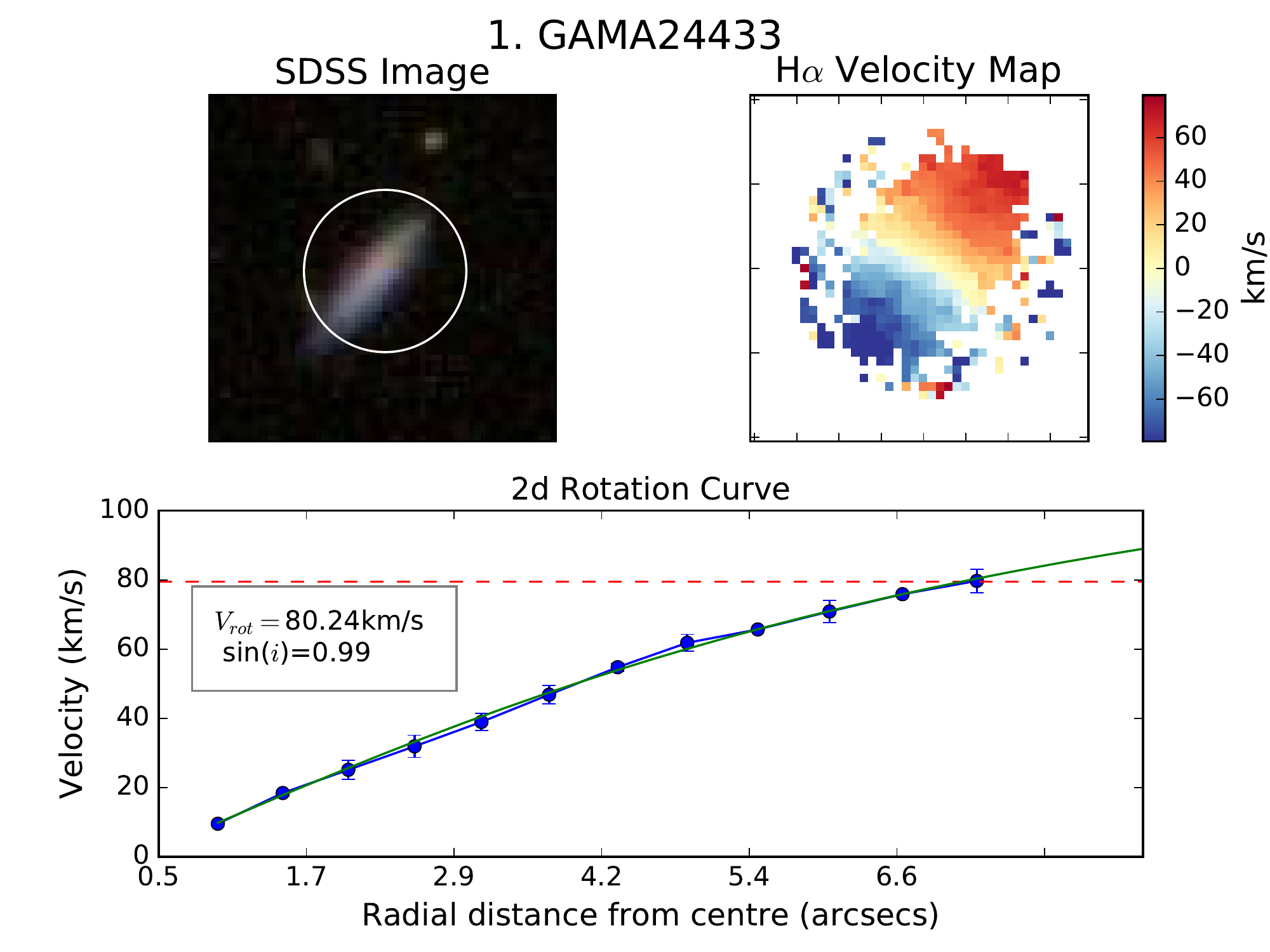}
\includegraphics[width=7cm]{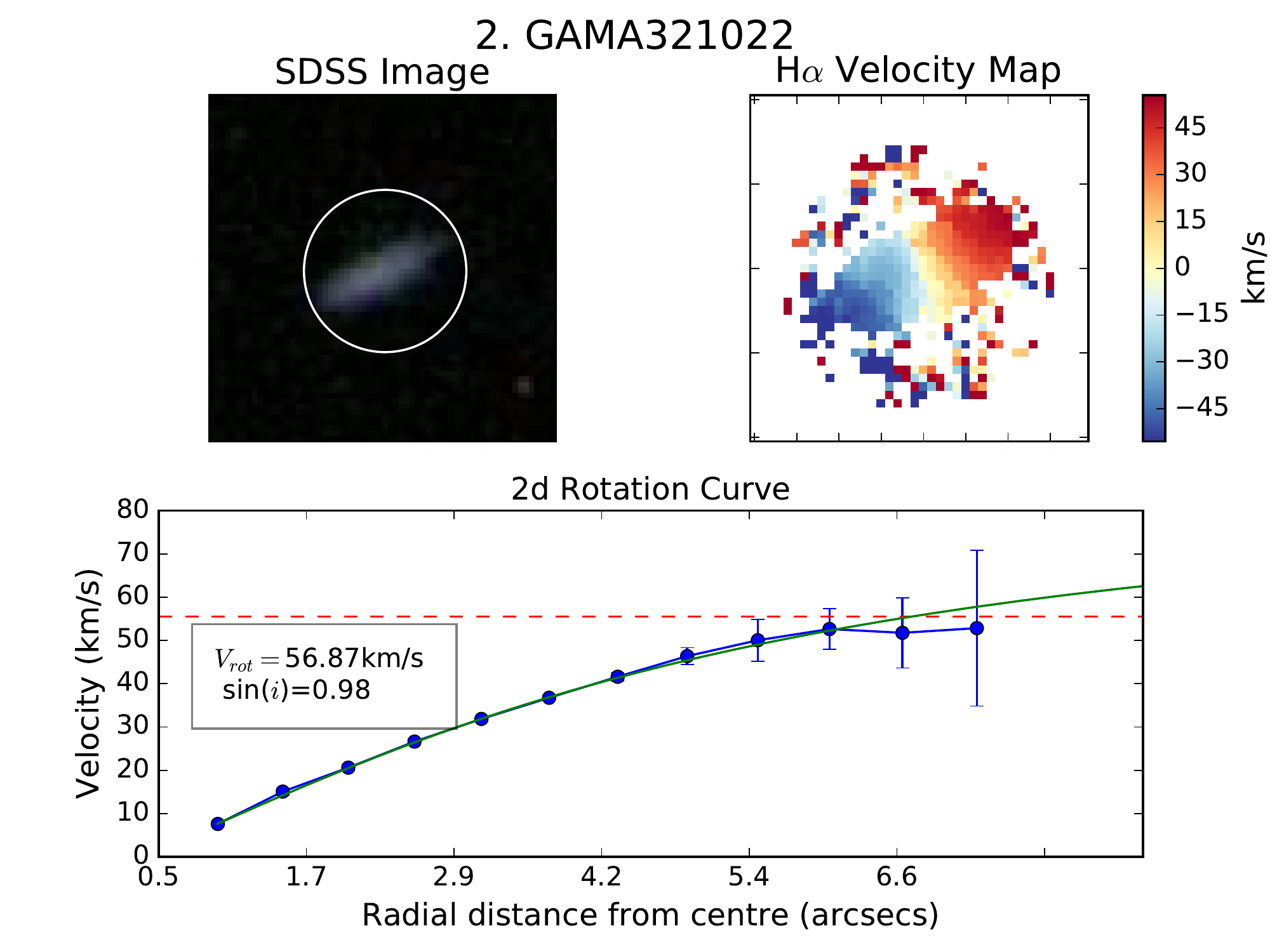}
\includegraphics[width=7cm]{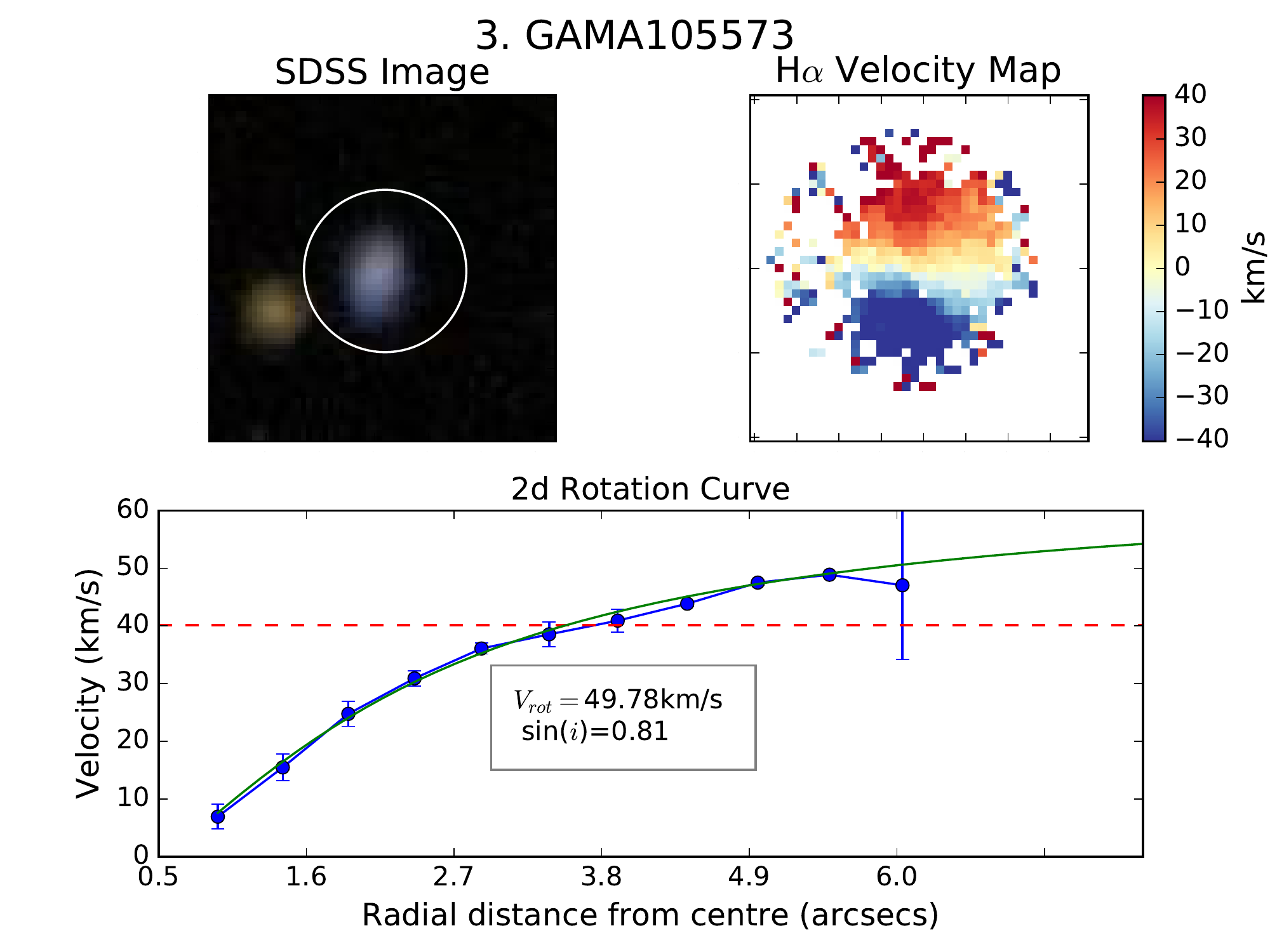}
\includegraphics[width=7cm]{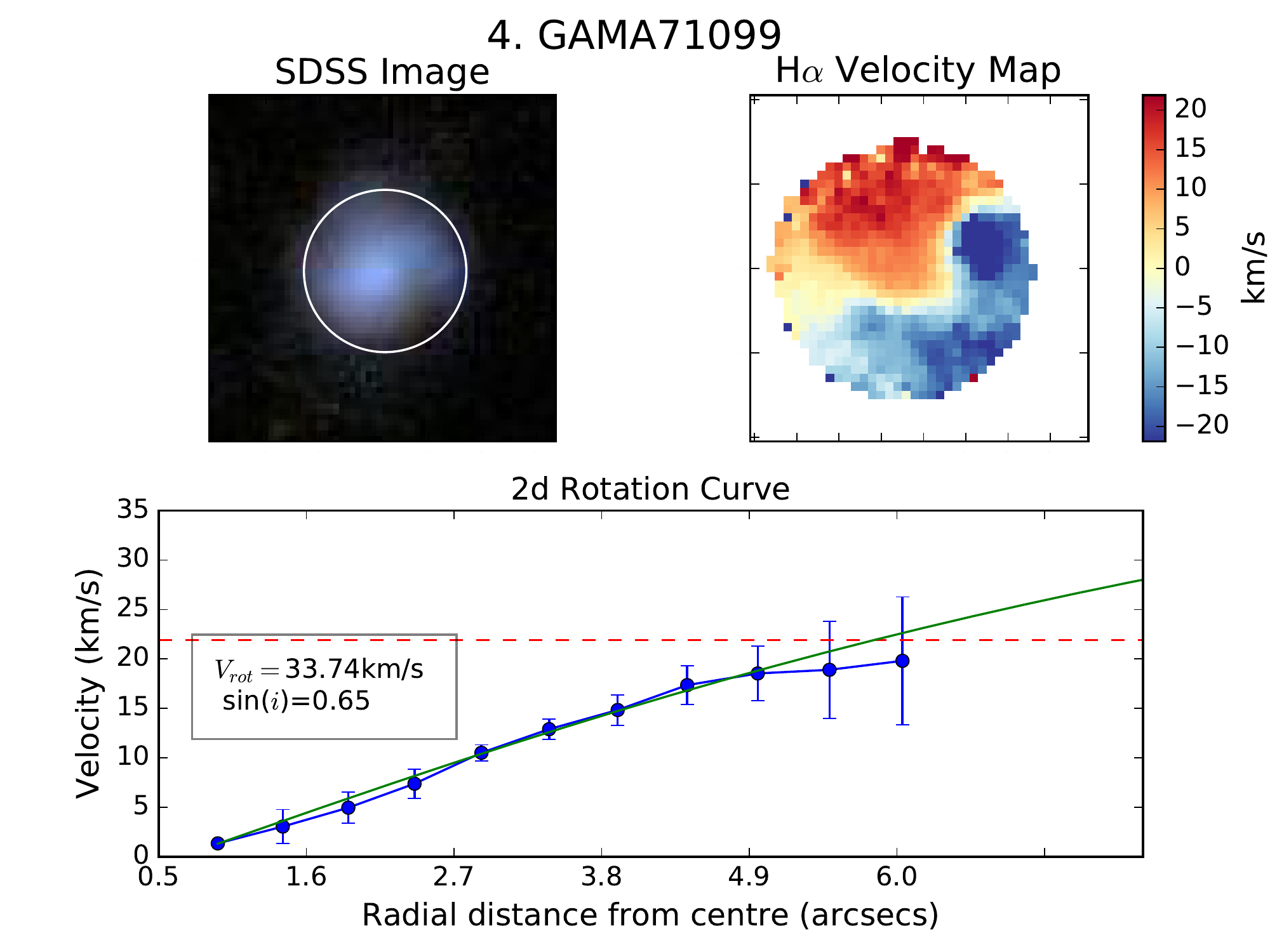}
\includegraphics[width=7cm]{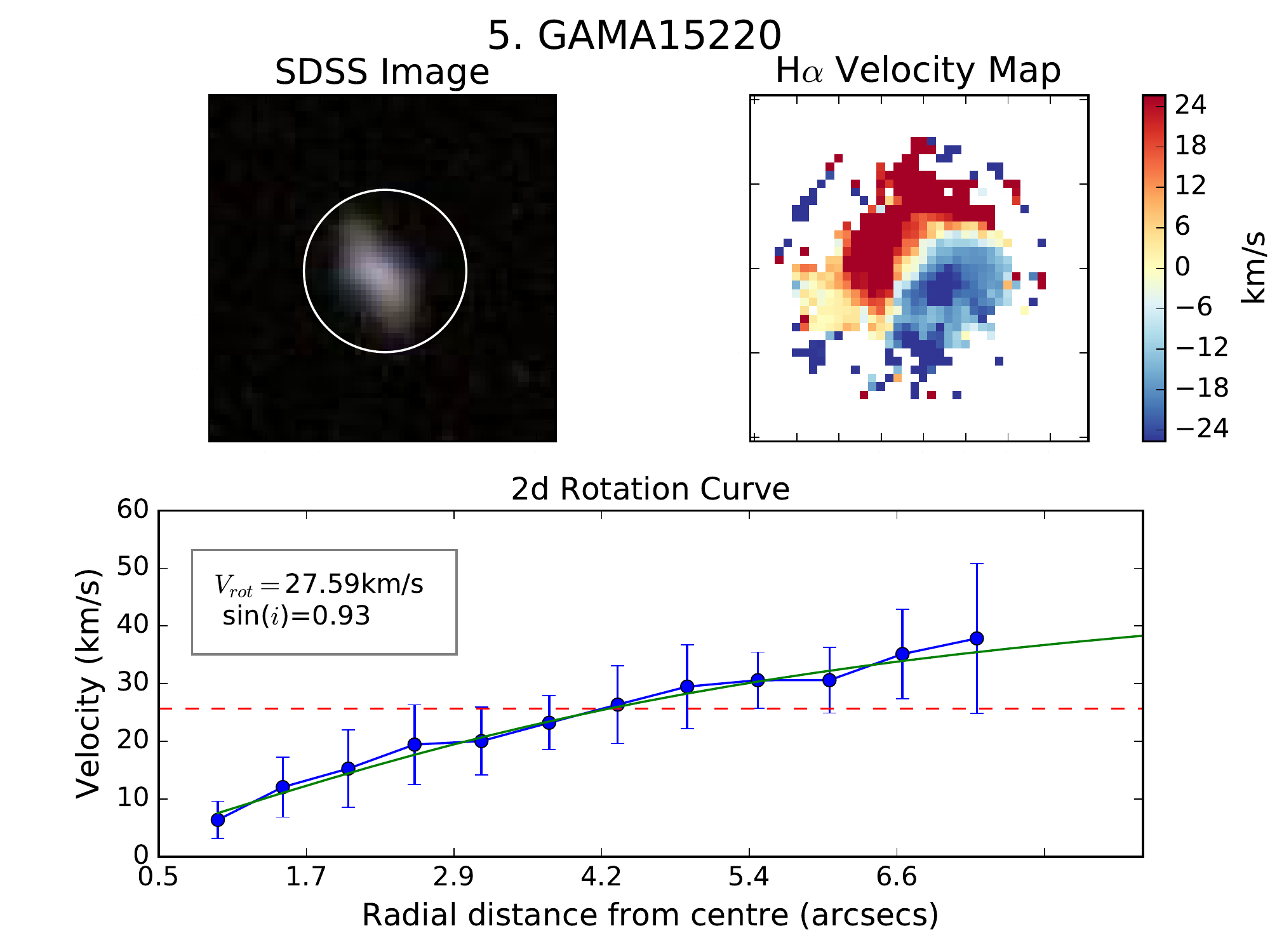}
\includegraphics[width=7cm]{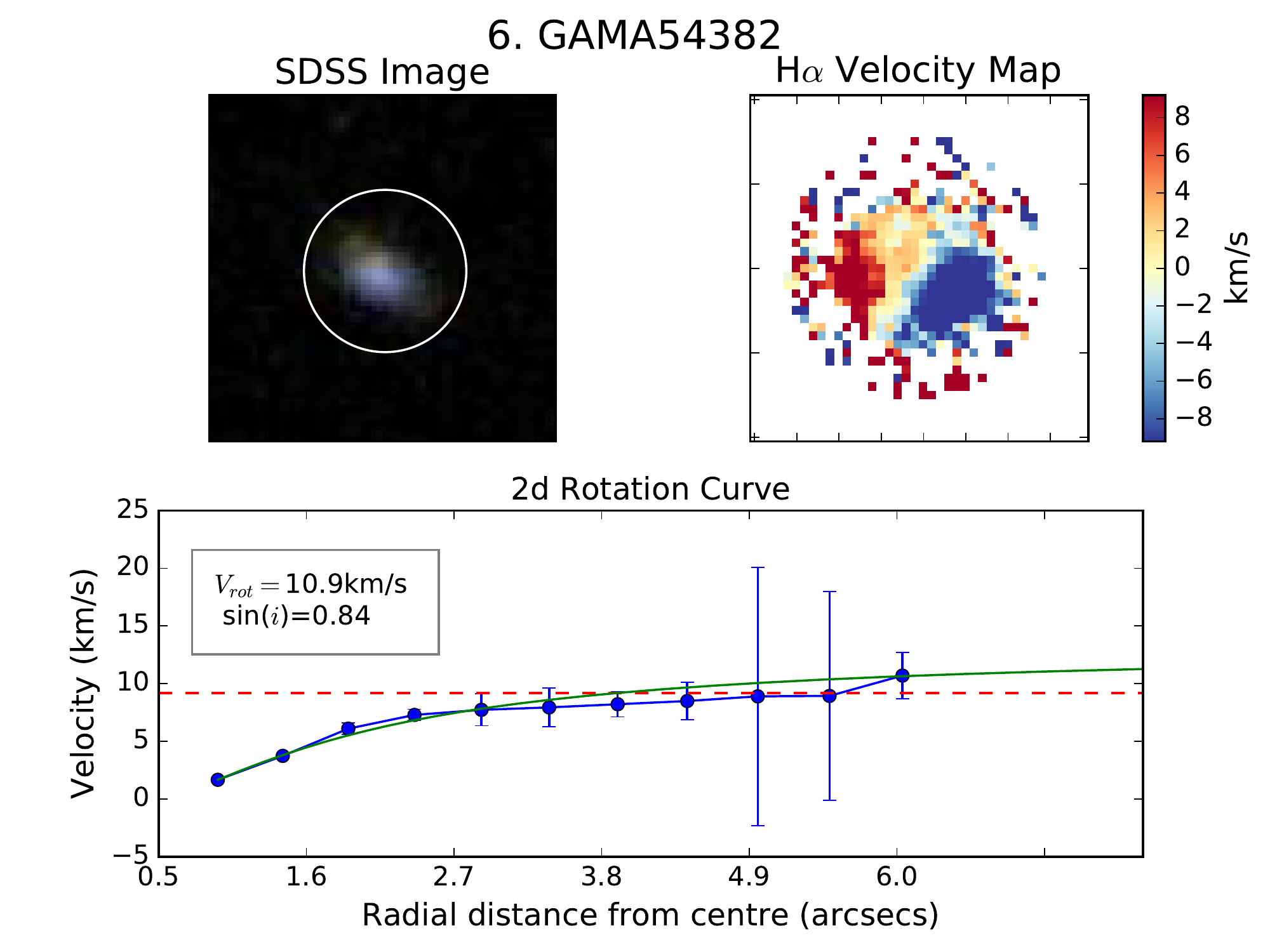}
\includegraphics[width=7cm]{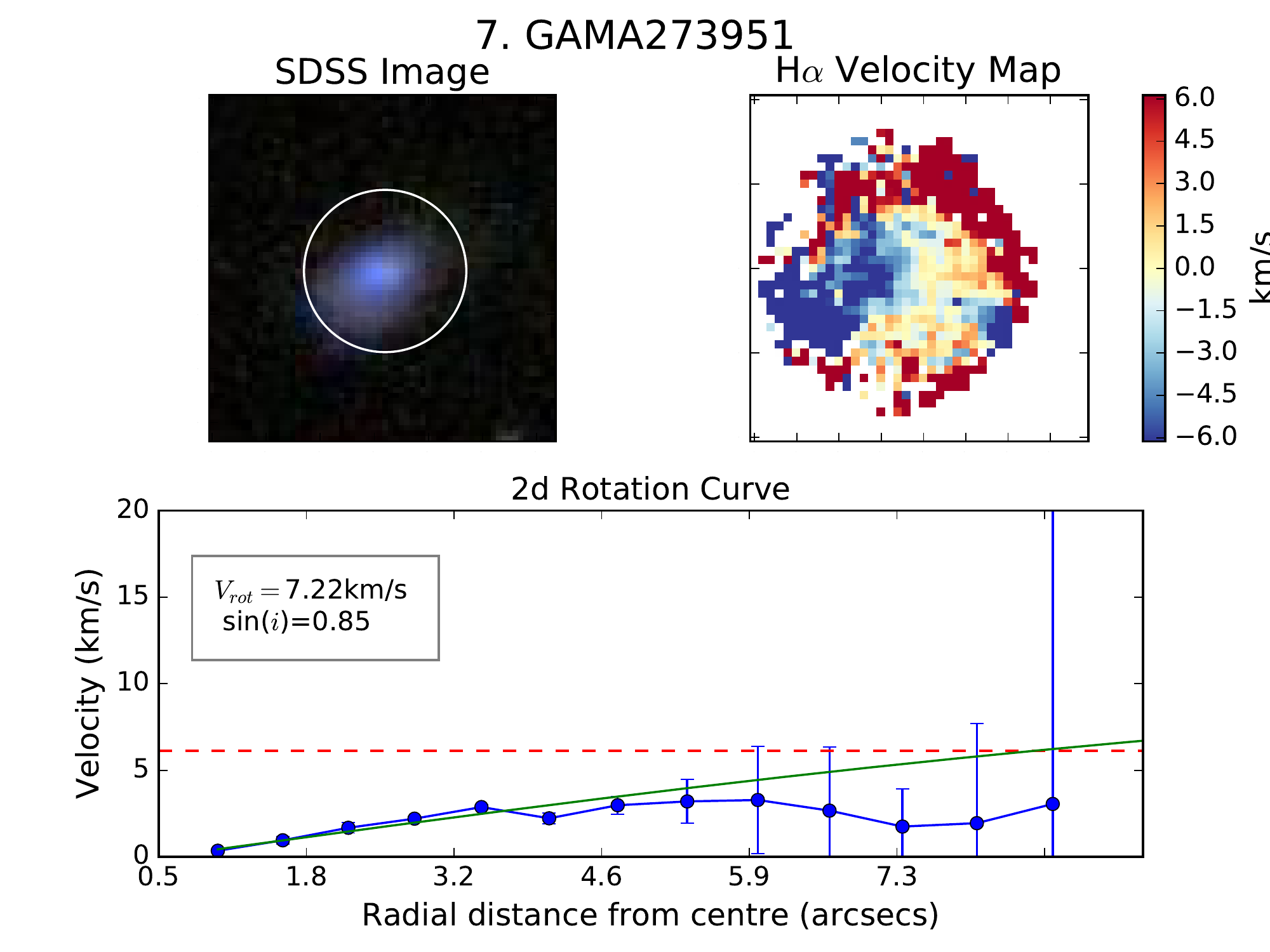}
\caption{The TFR as in Fig.~\ref{fig:twod_tfr_asyms}, with examples of low mass galaxies to illustrate that we are able to fit low stellar mass galaxies with a variety of velocities. The top left panel shows the TFR with the examples in colours. For each galaxy, the SDSS image is shown with the 15 arcsec SAMI field of view as a white circle (left), and the $H\alpha$ velocity field with S/N$>6$ (right). The $k1$ values from kinemetry are plotted, as well as the fitted rotation curve. The red horizontal line gives $V_{rot}$ at 2.2$r_e$, uncorrected for inclination. Corrected rotation velocities and inclination corrections are given within the figures. }
\label{fig:lowm_fits}
\end{figure*}

Fig.~\ref{fig:lowm_fits} shows examples of fits to low mass galaxies, to demonstrate that galaxies are fit well even at the low mass end of the TFR. {In some cases, there is patchy data with S/N$>6$ around the edges of the velocity field. This does not affects the fits, as kinemetric ellipses are constrained by PA and ellipticity, and are only included if $<75\%$ of points in the ellipse contain valid data. There may, however, be some uncertainty introduced by unreliable $r_e$ measurements, due to the difficulties inherent in performing photometry on small, low surface brightness galaxies. We discuss this more fully in Section~\ref{sec:errscat}. } 

Fig~\ref{fig:simons} shows our TFR in comparison with those from \citet{green2013dynamo} and \citet{simons2015transition} and our linear fit from Fig.~\ref{fig:twod_tfr_asyms} in comparison to that from \citet{reyes2011calibrated}. 
Our slope value is in agreement with the results from \citet{reyes2011calibrated}, who take a sample of 189 galaxies from the SDSS sample with $z<0.1$ over mass range $9.0<\log(M_*/M_{\odot})<11.0$ to find a slope of $0.278\pm0.13$ (see Fig.~\ref{fig:twod_tfr_asyms}). At high stellar mass, our TFR is in agreement with the results of \citet{green2013dynamo} who, similarly to this work, use 2D spatially resolved H$\alpha$ kinematics to fit rotation velocities at 2.2$r_e$ for low redshift galaxies.

{We also show the TFR line from \citet{bekeraite2016space}, who use IFS to calculate rotation velocity from stellar kinematic maps. Comparison to this work is directly relevant, due to the closeness of their redshift ($0.005 < z < 0.03$) and stellar mass ($9.0<\log(M_*/M_\odot)<12.0$) ranges to the SAMI Galaxy Survey. Their work shows SDSS Petrosian absolute magnitude ($M_r$) against circular velocity. In order to directly compare our measurements, we derive a conversion factor by taking the mean ratio of $M_r$ to stellar mass for our galaxies with $\overline{v_{asym}}<0.04$ and $\log(M_*/M_\odot)<9.0$. We use these cuts to exclude high asymmetry galaxies (as for producing Equation~\ref{equation:line}) and so as not to extend below the CALIFA mass range. The resulting converted CALIFA stellar mass TFR is given by:
\begin{equation}
\mathrm{log}(V_{rot}/kms^{-1})=0.26 \pm 0.017 \times \mathrm{log}(M_{*}/M_{\odot})- 0.50 \pm 0.13
\end{equation}
This line is shown in Fig.~\ref{fig:simons}. Despite the differences in method, the \citet{bekeraite2016space} fit agrees with the fit to our work within the errors. {We note that that our conversion factor was derived by taking a simple mean, which ignores differences in mass/light ratio for galaxies of different ages, which, if not ignored, may introduce a change in the derived slope.} {When comparing between different TFR studies, it is also important to note that many studies use different definitions of velocity. For example, \citet{bekeraite2016space} define rotation velocity at $r_{opt}$, the radius containing $83\%$ of the light. Different definitions of rotation velocity can lead to different TFR slopes if measured at a radius where the rotation curve has not flattened \citep{bradford2016slippery}. We note that at  $r_{opt}$, most galaxies' rotation curves will have flattened, and depending on structural morphology it is a similar radius to $2.2r_e$, which was selected for the same reason. }

\citet{simons2015transition} found a transition mass in the TFR at $\log(M_{*}/M_{\odot})\sim9.5$, below which galaxies were almost all scattered lower than the relation, in contrast to our results. Fig.~\ref{fig:res_hists} shows histograms of residuals around the fitted TFRs for each data set, in bins of stellar mass. In the $\mathrm{log}(M_{*}/M_{\odot})<9.0$ and $9.0<\mathrm{log}(M_{*}/M_{\odot})<10.0$ cases, there is an excess of negative residuals in the \citet{simons2015transition} distribution, in comparison to the distributions from the SAMI Galaxy Survey data, with two-sample Kolmogorov-Smirnov tests giving $p$-values of 0.018 and 0.023, respectively.  In both data sets, there is a progressive increase in negative scatter as stellar mass decreases. The discrepancy may have been caused by the difficulty in accurate slit placement for low mass, asymmetric galaxies. \citet{simons2015transition} themselves discuss the difficulty of accurate slit placement, and  given the higher proportion of asymmetric galaxies at low stellar mass seen in Fig.~\ref{fig:twod_tfr_asyms}, a number of the galaxies in \citet{simons2015transition} may have inaccurate values for $V_{rot}$. Finally, the \citet{simons2015transition} sample is at significantly higher redshift than ours ($\overline{z}=0.26$, rather than $\overline{z}=0.03$), leading to lower spatial resolution. {There may be redshift evolution between the \citet{simons2015transition} sample and our sample. However, the TFR slope at the high mass end is similar for both samples and therefore does not show the evolution suggested in \citet{cresci2009sins} and \citet{ziegler2001evolution}. Although change in TFR slope has been found in several studies of TFR evolution, there has been no work to date investigating scatter at the lower mass end as an effect of evolution of galaxy kinematics. The differences between slit and IFU data discussed above may mask any signatures of evolution of the TFR between \citet{simons2015transition} and our sample at low stellar mass. Comparable data types are needed to conduct a valid study of evolution, such as between the KROSS and SAMI samples in Tiley et al. (in prep.).}


In order to compare the low mass end of our TFR with other samples, we convert our stellar masses to baryonic masses. We approximate the total baryonic mass by adding the fractional gas mass to the stellar mass, but as we do not have direct gas masses for our sample, gas fractions are estimated from equation 5 in \citet{cortese2011effect}:
\begin{equation}
\log\bigg(\frac{M(HI)}{M_{*}/M_{\odot}}\bigg)=-0.33(NUV-r)-0.40log(\mu_{*})+3.37
\end{equation}
where $NUV-r$ is the {colour in magnitudes} and $\mu_{*}$ is the stellar mass surface density {in $\mathrm{M_*/pc^2}$}. The relationships between stellar mass, scatter and asymmetry observed in Fig.~\ref{fig:twod_tfr_asyms} remain in the baryonic mass TFR.

Fig.~\ref{fig:bar} shows our TFR with baryonic mass, instead of stellar mass. We compare with the HI rotation velocity and baryonic mass TFRs from the low-redshift, gas dominated ($M_{gas}>M_{star}$) sample in \citet{stark2009first}, {and the low mass compilation in \citet{papastergis2015big}, as discussed in \citet{sales2017low}}. These samples have selected against disturbed galaxies, so it is not surprising that our TFR shows more downwards scatter than theirs. However, it is notable that the main distributions are within a similar $V_{rot}$ range.

\begin{figure}
\centering
\includegraphics[width=9cm]{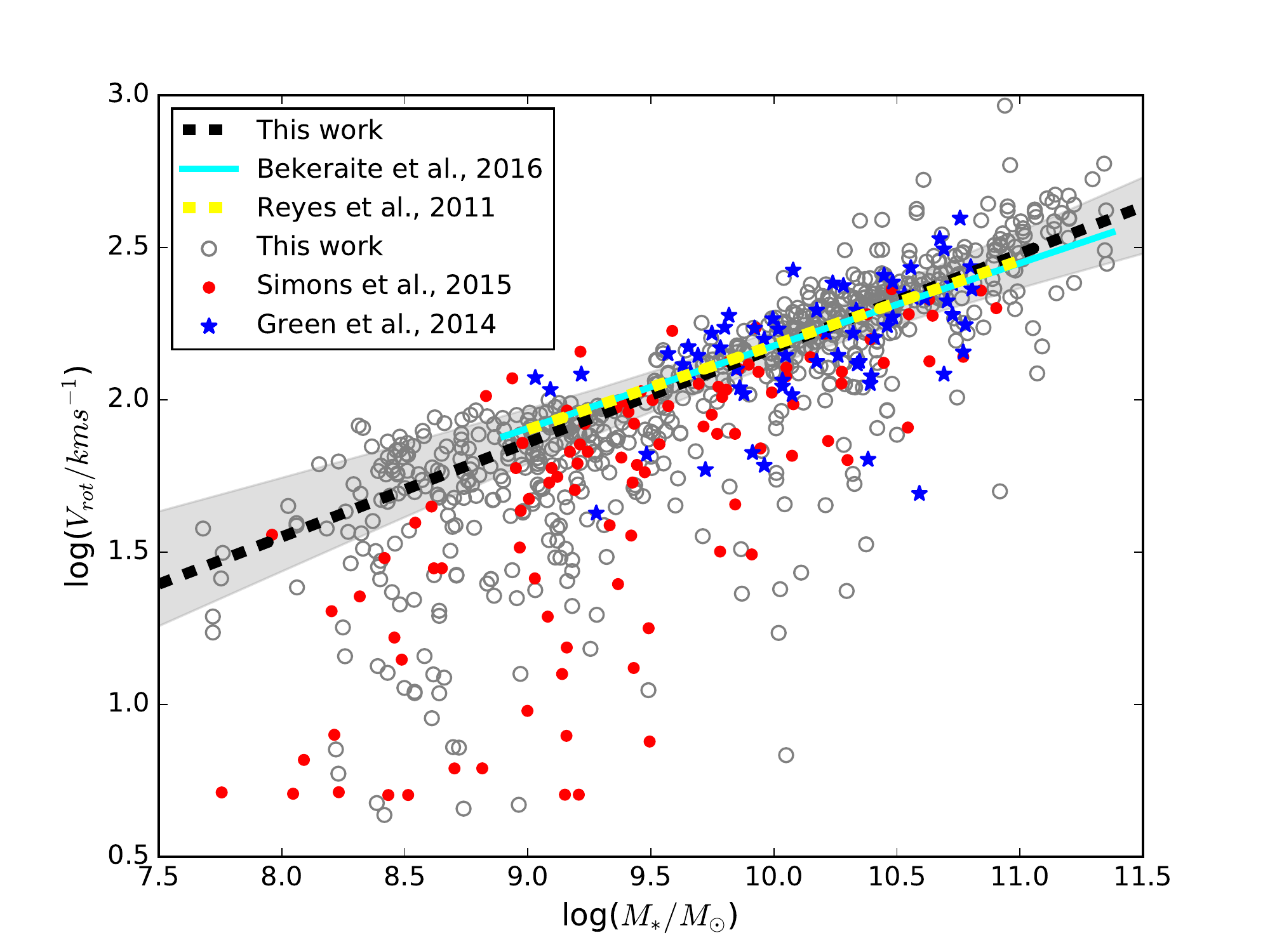}
\caption{The 2D TFR from this work shown with the stellar mass/H$\alpha$ TFR from \citet{green2013dynamo} and \citet{simons2015transition} and the fitted TFR lines from \citet{reyes2011calibrated} and \citet{bekeraite2016space}. The linear fit to our data from Fig.~\ref{fig:twod_tfr_asyms} is also shown. Our points and fit are consistent with the comparison samples at high mass. Below $\log(M_{*}/M_{\odot})\sim9.5$, however, there is disagreement between our results and those in \citet{simons2015transition}. Unlike \citet{simons2015transition}, we do not see a decay of the TFR below this cutoff, although there is an increase in negative scatter.}
\label{fig:simons}
\end{figure}

\begin{figure}
\centering
\includegraphics[width=9cm]{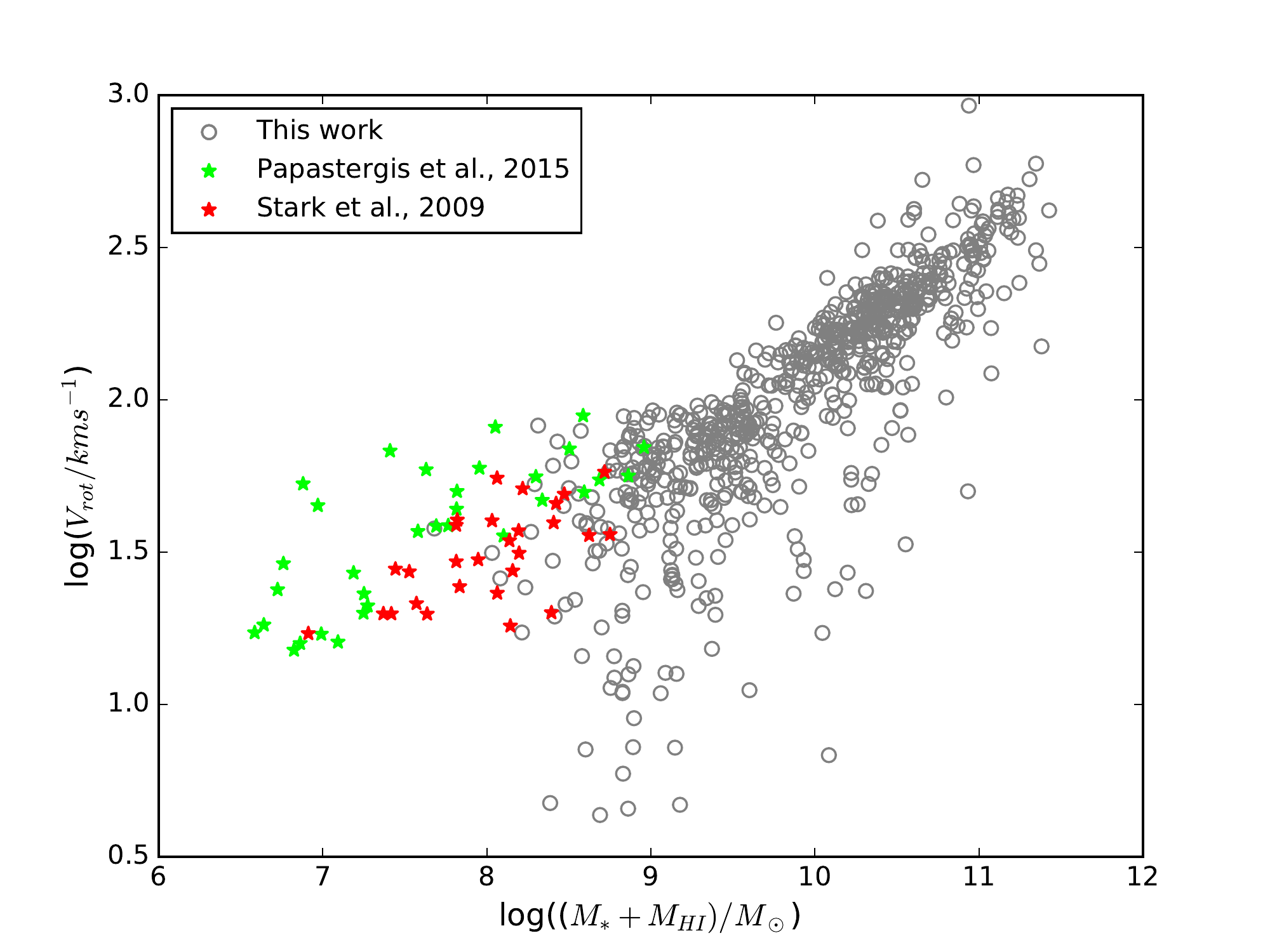}
\caption{ The 2D TFR with baryonic masses, shown with comparison samples from \citet{stark2009first} and the compilation in \citet{papastergis2015big}. These comparison samples follow a similar distribution to our work, taking into account that they selected against disturbed galaxies, and so show less scatter than our complete sample.}
\label{fig:bar}
\end{figure}


\begin{figure*}
\centering
\includegraphics[width=\textwidth]{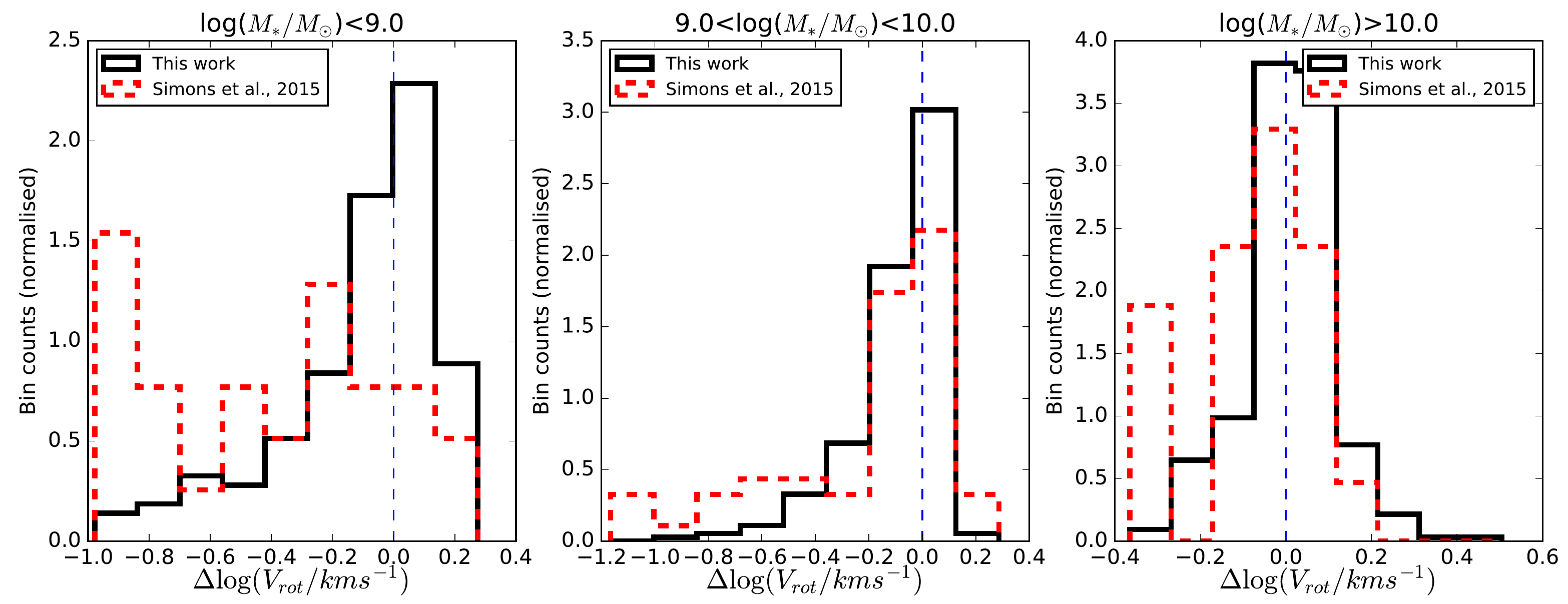}
\caption{Histograms of residuals around the TFR lines (shown in Fig.~\ref{fig:simons}) for the \citet{simons2015transition} data (red, dashed) and that in this work (black, solid), with the zero line in each case shown in dashed blue. In all cases, there is an excess of negative residuals in the \citet{simons2015transition} distribution, in comparison to this work. }
\label{fig:res_hists}
\end{figure*}

\subsection{`Simulated slit' TFRs}
\label{sec:simslits}

In order to examine the effects of slit measurement compared with 2D kinematics, we placed a `simulated slit' on the observed SAMI Galaxy Survey H$\alpha$ velocity fields. 

Using either photometric or kinematic position angles we selected spaxels from SAMI Galaxy Survey H$\alpha$ velocity maps along a `slit' of width 1 arcsec and the given PA. This slit width allows our results to be compared to other low-redshift TFR studies, including \citet{raychaudhury1997tests}, \citet{kannappan2002physical}, \citet{reyes2011calibrated} and others, that use slit widths 2 arcsec and smaller, and are comparable to our work on a physical scale. We note that slit widths of 2 arcsec or 3 arcsec do not alter the global scatter of the resulting TFR. We also select a narrow slit width in order to isolate the effects of changing the PA, and minimise spaxels selected in common when the PAs are offset. Velocity values for spaxels falling within the slit were plotted against radial distance from the fitted central spaxel, and arctan rotation curves were fit to the slit data, as above. We note that slit placement for our sample is not vulnerable to placement error, as is discussed in \citet{simons2015transition}, who consider acceptable slit placement to be within 40$^\circ$ of the photometric PA. Of course, some galaxies may have poorly defined PAs, such as in the case of strong bars (in the photometric case) or asymmetric galaxies (in the photometric and kinematic cases). However, once the PA is calculated, there is no error in placement.

Of the total parent sample of 813 galaxies, the highest fraction were fit when the input velocities were from 2D spatially resolved kinematics. Table~\ref{table:comb} gives fraction of galaxies fit for each method, and the number of galaxies not fit because they had $<5$ points with errors $<12$km/s, no turnover point or $\chi^2>200$, respectively. We applied the same cuts to the 2D and simulated slit samples. We choose a limit of $\chi^2=200$ because the distribution of $\chi^2$ {was strongly peaked at low values, and} for all three samples has a significant break between $200$ and $550$, making this a natural cutoff. {The reduced $\chi^2$ distribution also has a clear break, and separates out the same galaxies.} We note that no galaxies in the 2D spatially resolved sample were cut because they had $\chi^2>200$. In fact, 4 galaxies had $\chi^2>200$, but also had no turnover, which was the cut performed first. $82\%$ of galaxies in the 2D spatially resolved sample had $\chi^2<14$, and $90\%$ had $\chi^2<50$. Galaxies with higher $\chi^2$ were asymmetric galaxies with rotation curves deviating from the arctan model. However, the generally low $\chi^2$ values for fits to kinematically normal galaxies provide justification for the use of the arctan curve from  \citet{courteau1997optical}, as it is a good match for the rotation curves derived from the whole kinematic map. Typical degrees of freedom for each rotation curve are $\sim10$ for rotation curves from kinemetry and $\sim20$ for those from the simulated slit.




\begin{center}
\begin{table*}
  \begin{tabular}{ b{2.5cm}  l b{2cm}  l b{2cm}  l b{2.5cm}  l b{1.5cm}  l b{2.5cm}  l b{2.5cm}  } 
    \hline
     & Fraction fit & Too few points & No turnover & $\chi^2>200$ & \vline ~RMS (dex) & Median abs. residual from TFR (dex) \\ \hline
    2D Spat. resolved & $90\%$ & 59 & 21 & 0 &\vline ~0.15 & 0.058 \\
    Kin. PA slit & $74\%$ & 96 & 66 & 51 &\vline~ 0.17 & 0.063 \\
    Phot. PA slit & $68\%$ & 102 & 42 & 101&\vline~ 0.25 & 0.071 \\
    \hline
  \end{tabular}
\caption[Table caption text]{This table gives the fraction of galaxies fit for each method, and the number of galaxies not fit because they had $<5$ points with errors $<12$km/s, no turnover point or $\chi^2>200$, respectively. We also show global scatter measurements for the TFRs produced using our three methods (see Fig.~\ref{fig:common_tfrs}). Scatter (both RMS and median absolute residual from TFR as in Equation~\ref{equation:line}, in log space) is increased by using a slit, instead of the 2D spatially resolved map. Scatter is further increased when the slit is positioned at the photometric, rather than the kinematic PA.} 
\label{table:comb}
\end{table*}
\end{center}

\begin{center}
\begin{table*}
  \begin{tabular}{ b{2.5cm}  l  b{2.5cm}  l b{2.5cm} l b{2cm}}
 & \multicolumn{2}{c}{$\Delta V_{rot}$/Stellar Mass correlation} & \multicolumn{2}{c}{$\Delta V_{rot}$/$\overline{v_{asym}}$ correlation}& \multicolumn{2}{c} {TFR fit parameters} \\
    \hline
     & $\rho_{SM}$ & p-value$_{SM}$ &\vline ~ $\rho_{asym}$& p-value$_{asym}$&\vline ~ Slope & $y$-int.  \\ \hline
    2D Spat. resolved & $ -0.28$ & $2\times10^{-6}$&\vline ~  $ -0.30$ & $1.4\times10^{-8}$& \vline ~0.31 & -0.93\\
    Kin. PA slit & $-0.25$ & $4\times10^{-5}$ &\vline ~ $-0.32$ & $1.2\times10^{-9}$ &\vline ~0.30 & -0.84 \\
    Phot. PA slit & $-0.36$ & $2\times10^{-3}$ &\vline ~ $-0.25$ & $2.5\times10^{-6}$ & \vline ~0.29 & -0.83 \\
    \hline
  \end{tabular}
\caption[Table caption text]{This table gives Spearman rank correlation parameters (subscript $SM$) between scatter off the TFR and stellar mass for the three TFRs in Fig.~\ref{fig:common_tfrs}. In all three cases, scatter below the TFR is inversely correlated with stellar mass. We also give correlation parameters (subscript $asym$) for $\overline{v_{asym}}$ and scatter off the TFR for the three samples in  Fig.~\ref{fig:common_tfrs}. The relationship holds in all three cases. Note that the correlations are negative because the main scatter is below the TFR. {Finally, we give TFR fit parameters for all three samples, as shown in Fig.~\ref{fig:common_tfrs}.}} 
\label{table:tabscat}
\end{table*}
\end{center}



The reason for the larger number of fitted galaxies from 2D kinematics is that the $k1$ values used in the 2D spatially resolved case represent bulk rotation, without asymmetry. Further, using the 2D spatially resolved data allows for more spaxels to be included, reducing the proportional influence of individual, noisy data points or fluctuation in the velocity field. We note that there is no systematic difference in the kinematic asymmetry of the three samples, with the median asymmetries for the 2D spatially resolved maps, kinematic and photometric slits being $0.040\pm0.002$ $0.038\pm0.002$ and $0.039\pm0.002$, respectively. Typical uncertainties are also equivalent.

Fig.~\ref{fig:normal} shows example plots for a normal galaxy, with low kinematic asymmetry and aligned photometric and kinematic PAs. Fig.~\ref{fig:pa}, by contrast, shows a kinematically normal galaxy for which there is significant misalignment between the photometric and kinematic PA. This misalignment causes a $\sim40$km/s difference in calculated rotation velocity when using slits placed at the different PAs and illustrates how using the photometric PA to measure rotation velocity can increase scatter off the TFR. Fig.~\ref{fig:asym} shows an asymmetric, low mass galaxy, for which the PAs are also misaligned. In this case, asymmetry makes determination of an accurate PA in either case difficult. However, the comparative smoothness of the rotation curve generated from the 2D spatially resolved map, compared to that from the slits, demonstrates the value of the map method in producing a better rotation curve fit. We note that the slit rotation curves are not always centred on 0. This is because the slit is centred on the photometric centre of the galaxy, which can be offset from the kinematic centre of the map. {Because the ellipses used to generate the rotation curves from kinemetry use the photometric, rather than the kinematic, centre, the curves do not always go through $(0,0)$, as there are often slight offsets between the centres of the flux and H$\alpha$ velocity maps. Given that the offset is, in all cases, less than the error on the fitted velocity, we do not consider this a significant cause of scatter. {We also note that the offsets are not proportional to $\overline{v_{asym}}$, making it unlikely that small differences between the photometric and kinematic centres contribute significantly to asymmetry.}}
\begin{figure*}
\centering
\includegraphics{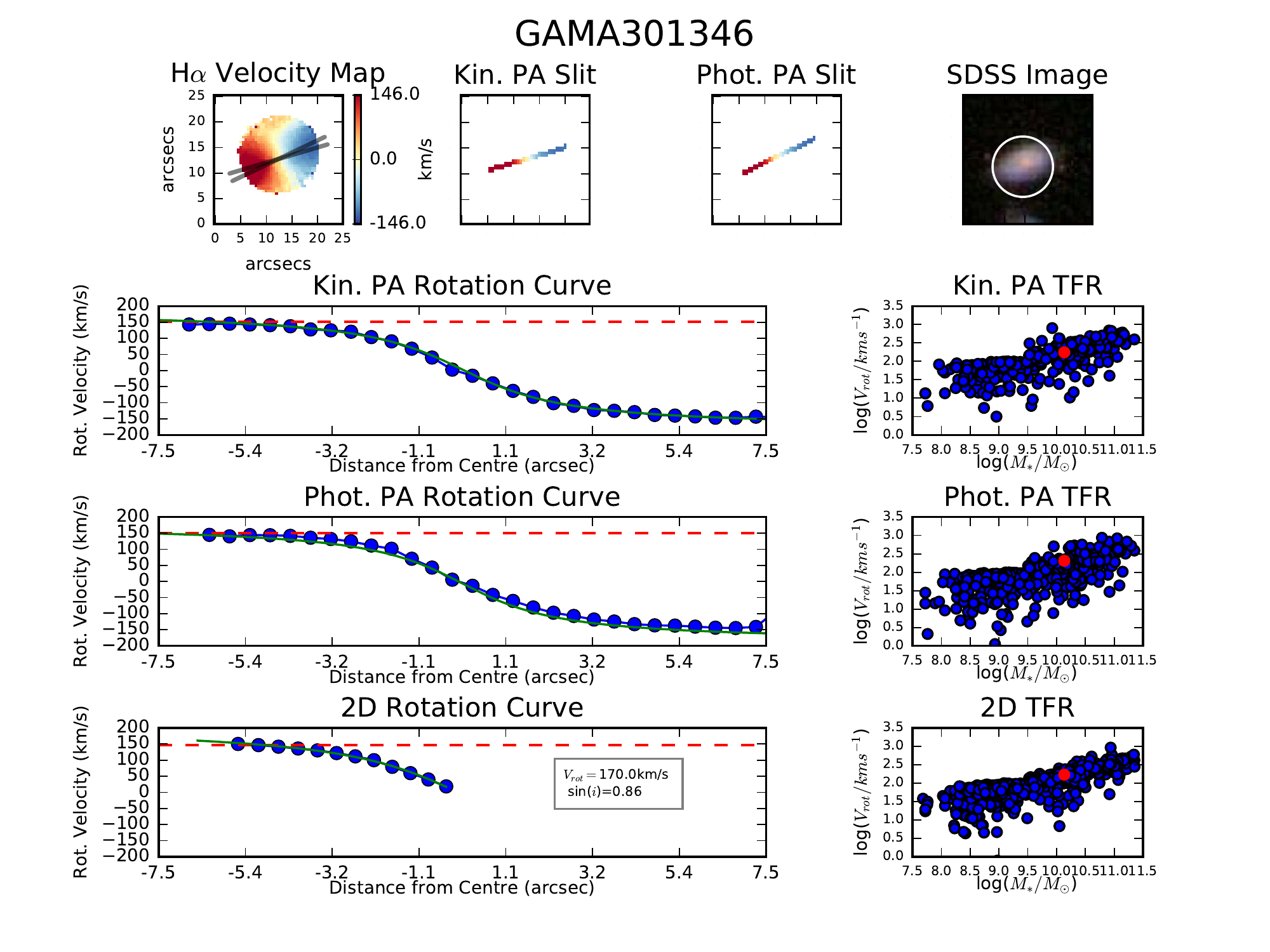}
\caption{We show fit results for GAMA310346, as an example of a kinematically normal galaxy, in which the photometric and kinematic PAs are aligned. The top row shows: the H$\alpha$ velocity map from {\small LZIFU}{\sc} (with the kinematic and photometric PAs overplotted) cut by S/N, the simulated slit spaxels at the kinematic and photometric PAs and the SDSS image. The SAMI instrument field of view is shown as a white circle. We then show the fitted rotation curve (green) and position on the TFR (red) for the simulated slits and 2D spatially resolved maps, respectively. On the plots of the rotation curves, the red dashed line shows the velocity (without inclination correction) at 2.2$r_e$. {The 2D rotation curve is a symmetric one-sided rotation curve, which has been plotted on the same axis scale as the others for ease of comparison.}}
\label{fig:normal}
\end{figure*}

\begin{figure*}
\centering
\includegraphics{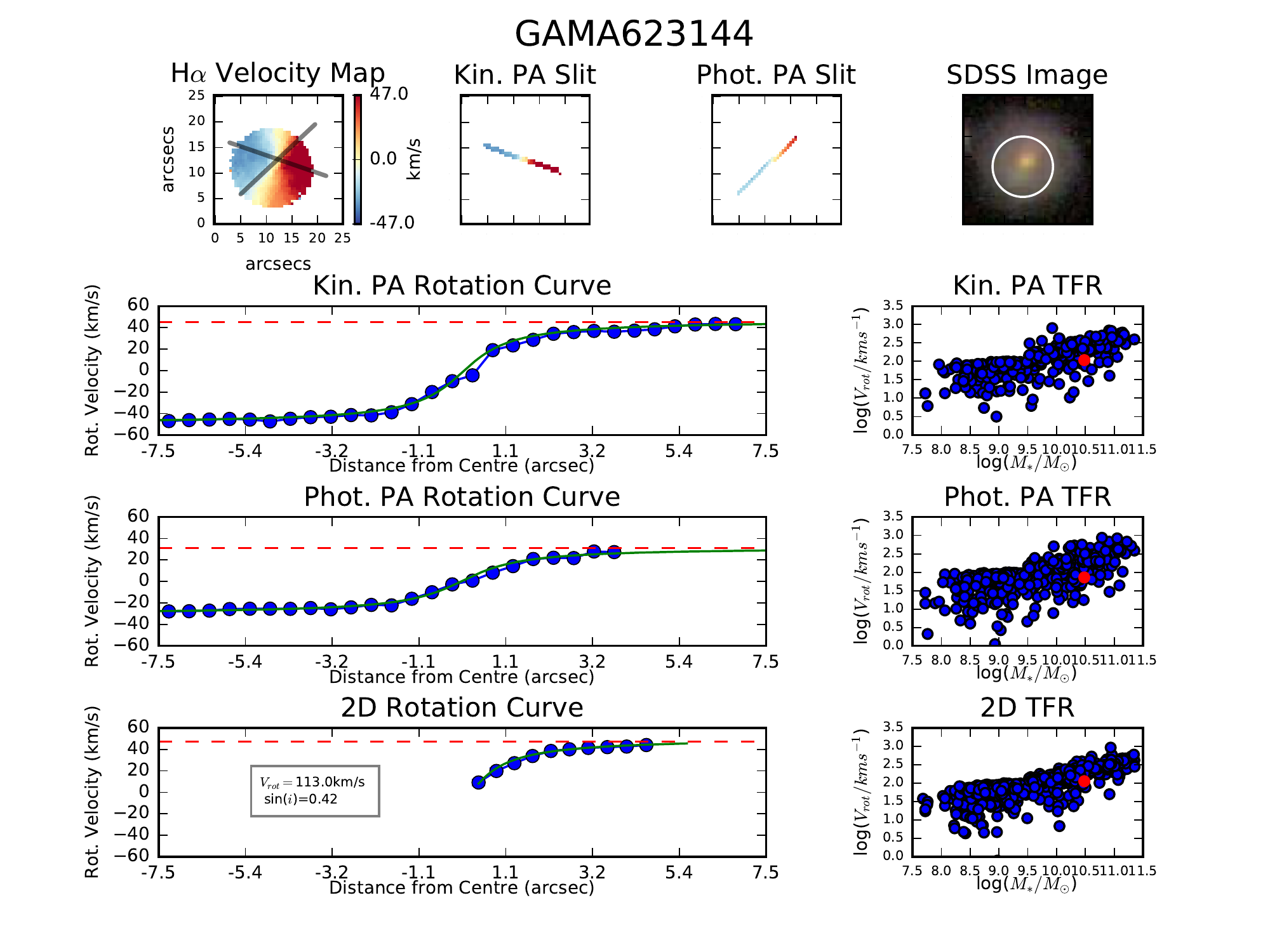}
\caption{A galaxy with offset photometric and kinematic PAs. The circular velocity using the 2D spatially resolved map is 113.0 km/s, similar to the value calculated when using the slit at the kinematic PA (115.2km/s). However, when the slit is placed at the photometric PA, the circular velocity is found to be 77.0 km/s. Note that this galaxy is not centred in the SAMI instrument field of view.}
\label{fig:pa}
\end{figure*}

\begin{figure*}
\centering
\includegraphics{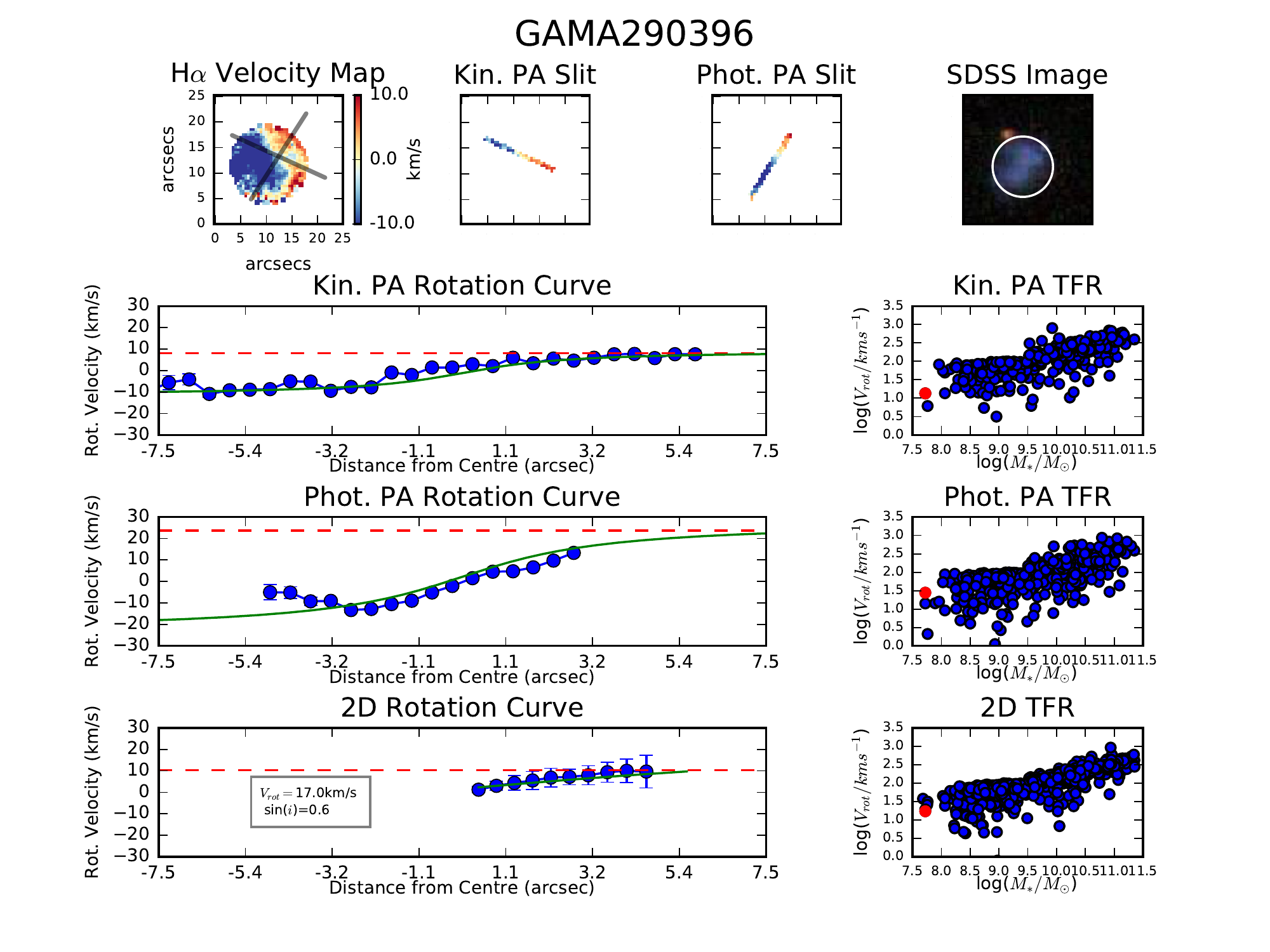}
\caption{A highly asymmetric ($\overline{v_{asym}}=0.17$) galaxy at low stellar mass. The photometric and kinematic PAs are significantly offset (although, given the level of disturbance, calculating an accurate PA is difficult). The rotation curve from the 2D spatially resolved map is much smoother than those from the simulated slits. This galaxy is also not centred in the SAMI instrument field of view.}
\label{fig:asym}
\end{figure*}

Fig.~\ref{fig:kinpa_slit_tfr} and Fig.~\ref{fig:photpa_slit_tfr} show the TFR produced using a slit placed along the photometric and kinematic PA, respectively.

\begin{figure}
\centering
\includegraphics[width=9cm]{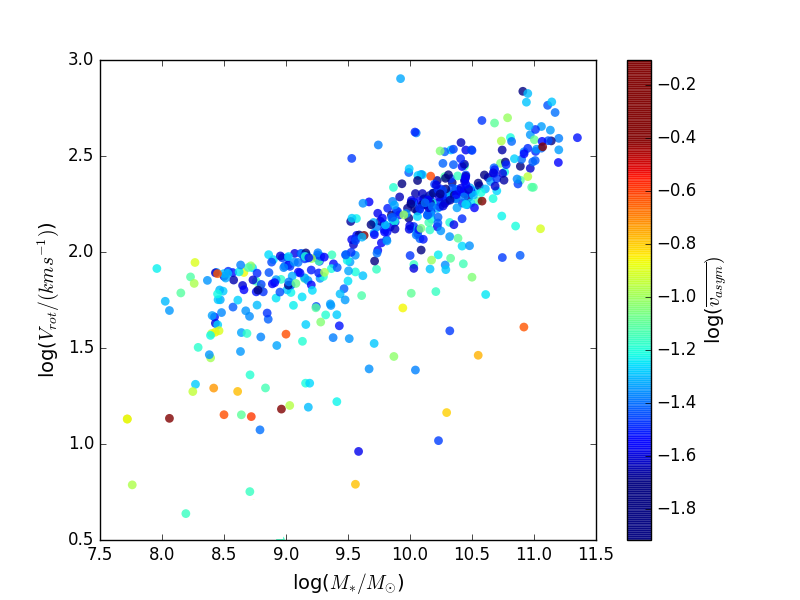}
\caption{The TFR with log($M_{*}/M_{\odot}$) against log($V_{rot}/km s^{-1}$) at 2.2$r_e$, calculated using rotation values from a slit placed along the kinematic PA. Points are coloured by $\log(\overline{v_{asym}})$. The only change between this plot and Fig.~\ref{fig:photpa_slit_tfr} is the angle at which the slit is placed. }
\label{fig:kinpa_slit_tfr}
\end{figure}

\begin{figure}
\centering
\includegraphics[width=9cm]{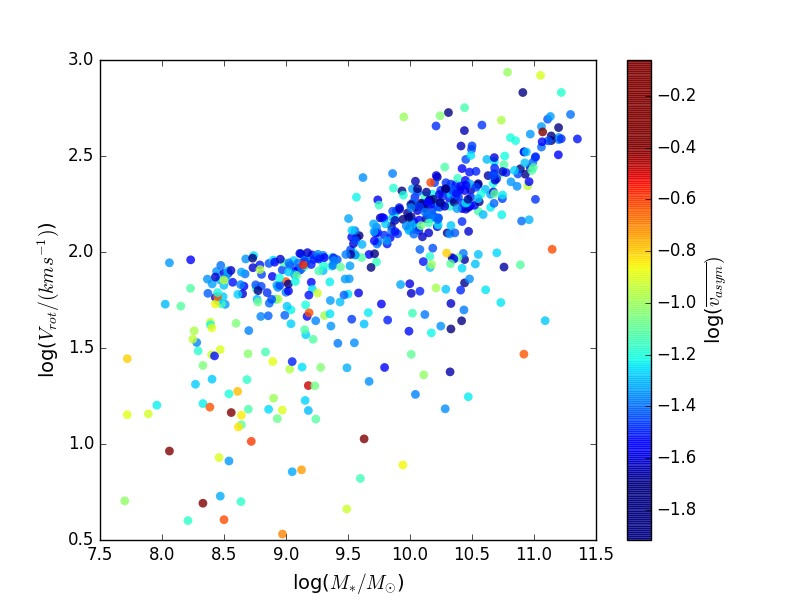}
\caption{The TFR with log($M_{*}/M_{\odot}$) against log($V_{rot}/km s^{-1}$) at 2.2$r_e$, calculated using rotation values from a slit placed along the photometric PA. Points are coloured by $\log(\overline{v_{asym}})$. There is more scatter than in Fig.~\ref{fig:twod_tfr_asyms}, particularly at low stellar mass. The main trends of Fig.~\ref{fig:twod_tfr_asyms}, however, are still visible: the TFR extending to low stellar mass and a correlation between scatter and $\overline{v_{asym}}$.}
\label{fig:photpa_slit_tfr}
\end{figure}

\section{Causes of scatter}
\label{sec:scatter}
Scatter off the TFR can be caused either by observational effects or by physical properties of galaxies. Understanding the scatter is thus crucial, both to more accurately plot the TFR and to understand why galaxies scatter. Processes known to correlate with scatter, such as asymmetry \citep{kannappan2002physical}, are also important for understanding galaxy evolution. 

Fig.~\ref{fig:common_tfrs} shows the TFRs produced by the three methods (2D kinematics, a slit at the kinematic PA and one at the photometric PA), including only the 456 galaxies common to all three. Table~\ref{table:comb} gives the RMS and median absolute residuals off the TFR line for the three TFRs in Fig.~\ref{fig:common_tfrs}. There are a variety of causes of scatter, which affect the TFRs in different ways:
\begin{itemize}
\item{Photometric uncertainties}
\item{Kinematic asymmetry}
\item{Scatter caused by using a slit}
\item{Underestimation of $V_{rot}$ caused by not measuring velocity along the kinematic PA}
\end{itemize}

We discuss these effects in turn in the following sections.

\begin{figure*}
\centering
\includegraphics[width=\textwidth]{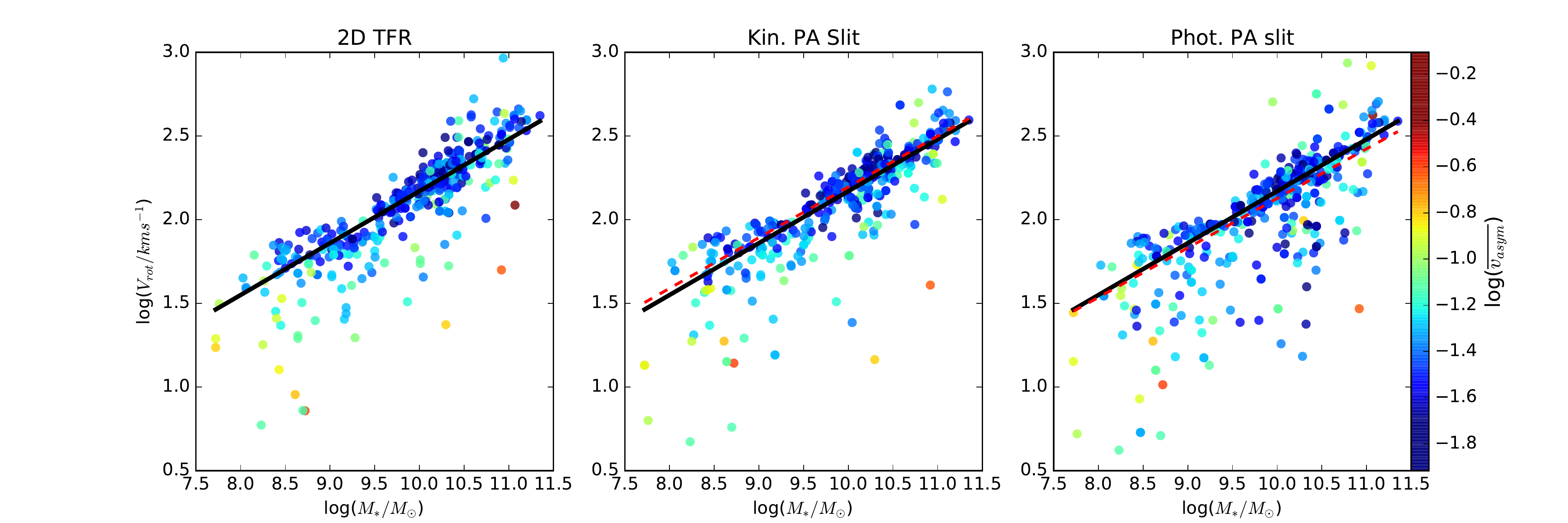}
\caption{TFRs produced using three methods, showing only galaxies common to all three. Points are coloured by kinematic asymmetry, and the black line shown gives the fitted TFR from Fig.~\ref{fig:twod_tfr_asyms}. The decrease in scatter from left to right is demonstrative of the influence of PA and the kinemetry method of moment extraction. {The red dashed lines show the TFR fit line for the galaxies common to all three using the kinematic and photometric PA, respectively. Slopes and intercepts are given in Table~\ref{table:tabscat}.}}.
\label{fig:common_tfrs}
\end{figure*}


\subsection{Photometric uncertainties}
\label{sec:errscat}
Small, low surface brightness galaxies may have less reliable photometry than large, bright galaxies, as has been noticed in \citet{hill2011galaxy} and \citet{simons2015transition}. The typical error value for MGE fits is 0.05 on $\log(r_e)$, but it is anticipated that there may be small increases in error on PA and ellipticity at low stellar masses. In order to quantify the photometric uncertainty, we calculated the difference in PA, $r_e$ and ellipticity from MGE photometry and the GAMA Sersic catalogue \citep{kelvin2012galaxy}. The RMS of the differences in PA, $r_e$ {(normalised to the MGE $r_e$)} and ellipticity are $14.6^{\circ}$, {0.24} and 0.12, respectively. The fractions of galaxies with scatter $>3\sigma$ are $3\%, 2\%$ and $1\%$, respectively. {We note that we expect the slightly larger scatter and RMS values for difference in the PA because of the large uncertainties associated with calculating a PA for very circular galaxies. Nevertheless, {the fact that} the scatter and RMS remain small in all cases indicates that the Sersic and MGE values are largely equivalent. }

It is possible that a small amount of scatter at low stellar mass is enhanced by errors in photometry. {However, causes of scatter are discussed below and, in particular, errors in stellar mass are small even at low stellar mass, averaging $0.1$dex, and are thus insufficient to explain the large scatter that we observe.}


\subsection{Kinematic asymmetry}

In all cases in Fig.~\ref{fig:common_tfrs}, there remains the correlation between kinematic asymmetry and scatter below the TFR, as in Fig.~\ref{fig:twod_tfr_asyms} and Fig.~\ref{fig:p_a_s}. Table~\ref{table:tabscat} gives correlation parameters for this relationship in all three cases.

There is also an inverse relationship between stellar mass and scatter off the TFR. Using the linear fit TFR from Fig.~\ref{fig:twod_tfr_asyms}, we calculate the negative scatter correlation with stellar mass and present the results for the common galaxies in Table~\ref{table:tabscat}. 

Given the known inverse relationship between stellar mass and kinematic asymmetry \citep{van1998neutral,cannon2004complex,lelli2014dynamics,bloom2016sami} and the relationships shown here between scatter off the TFR and kinematic asymmetry, this is not surprising. However, that the degree to which scatter depends on stellar mass changes between the presented TFRs (see Table~\ref{table:tabscat}) is indicative that it is more than a purely physical relationship, and could be at least partly a function of measurement. \citet{oh2016sami} demonstrated an inverse relationship between asymmetry and rotation velocity from stellar kinematics.

{The $S_{0.5}$ relation gives $\log(M_*/M_\odot)$ against $\sqrt{0.5V_{rot}^2+\sigma^2}$, where $\sigma$ is the mean velocity dispersion.} {It has been observed that asymmetric galaxies scatter less from the $S_{0.5}$ relation than they do from the TFR \citep{cortese2014sami,simons2015transition}. This indicates that asymmetry is related to the extent to which systems are dispersion, rather than rotation, supported. We shall investigate this further in future work.}


\subsection{Scatter caused by using a slit}

The slightly decreased scatter in the TFR from 2D kinematics compared with the kinematic slit TFR (see Table~\ref{table:comb}), is due to the derivation of the $V_{rot}$ values in the 2D kinematics TFR from the $k1$ moment map, that contains bulk rotation without asymmetry. $k1$ shows underlying disk rotation, calculated over all spaxels in the input velocity map, specifically excluding asymmetry and perturbation. By contrast, the velocity values used to produce the $V_{rot}$ values in the kinematic slit TFR are taken directly from the total velocity map, that includes asymmetries and fluctuations between spaxels. This can be seen in the difference between rotation curves from kinemetry and the kinematic PA slit in Figs.~\ref{fig:pa} and \ref{fig:asym}, in which the rotation curve from the kinematic slit is significantly less smooth than that from 2D kinematics.

Fig.~\ref{fig:vsvsvs_2D} shows $\log(V_{rot}/kms^{-1})$ from the kinematic PA slit against $\log(V_{rot}/kms^{-1})$ from the 2D kinematic maps, with points coloured by asymmetry.  A Spearman rank correlation test of asymmetry and absolute difference between $\log(V_{rot}/kms^{-1})$ from the kinematic PA slit against $\log(V_{rot}/kms^{-1})$ from the full kinematic maps gives $\rho=0.2, p=3\times10^{-5}$. The relationship between the $V_{rot}$ values is tight at high velocity, but below  $\log(V_{rot, 2D}/kms^{-1})\sim1.7$ the relation becomes much less tight. This indicates that the effects described above disproportionately affect asymmetric (and, by extension, low mass) galaxies.

The inherent lack of smoothing entailed by using a slit also contributes some of the increased scatter in the photometric slit TFR (see Table~\ref{table:comb}), as the rotation curves are produced in the same way as the kinematic slit TFR.

\begin{figure}
\centering
\includegraphics[width=9cm]{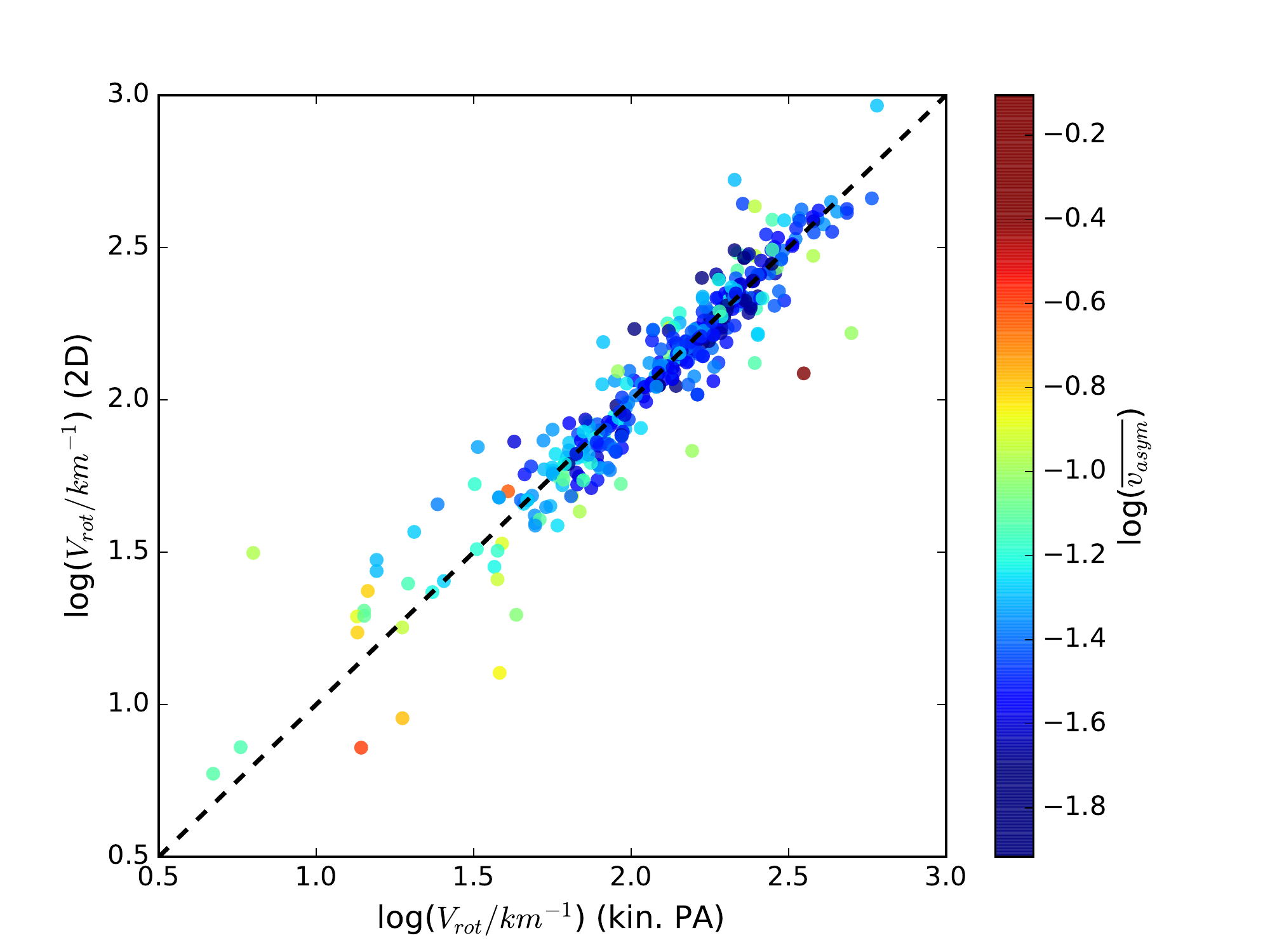}
\caption{$\log(V_{rot}/kms^{-1})$ from the kinematic PA slit against $\log(V_{rot}/kms^{-1})$ from the full kinematic maps, with points coloured by asymmetry. Difference between the measured $V_{rot}$ values is correlated with asymmetry, with an otherwise tight relation showing increased scatter below $\log(V_{rot, 2D}/kms^{-1})\sim1.7$. }
\label{fig:vsvsvs_2D}
\end{figure}

\subsection{Offset between photometric and kinematic PA}
\label{sec:offpa}

Other than input PA, there is no difference in velocity fitting or input data between the photometric and kinematic PA slit TFRs. From the increased scatter in the photometric slit, compared to the kinematic slit, TFR we conclude that scatter in the photometric slit TFR can be at least partially attributed to the fact that $23\%$ of galaxies common to all three TFR samples have misalignment $>30^{\circ}$. This misalignment leads to systematic underestimation of the rotation velocity when calculated along the photometric PA. In extreme cases, the photometric PA may lie along the zero line of the velocity map, entirely `missing' the galaxy's rotation (see Figs.~\ref{fig:normal}, \ref{fig:pa}, \ref{fig:asym}). Fig.~\ref{fig:vsvsvs} shows the underestimation of $V_{rot}$ for some galaxies produced by using the photometric, rather than kinematic PA.  There are multiple causes of kinematic and photometric misalignment. Table~\ref{table:paclass} gives the results of a by-eye classification of misaligned galaxies, done by examining rotation maps and SDSS images of misaligned galaxies. Of the misaligned galaxies, those with bars and those which are face-on can be considered to be misaligned due to observational, rather than physical effects. That is, the bar leads to the measured photometry being significantly different from the kinematics. If they were excluded from the photometric PA TFR, scatter would decrease.

\begin{figure}
\centering
\includegraphics[width=9cm]{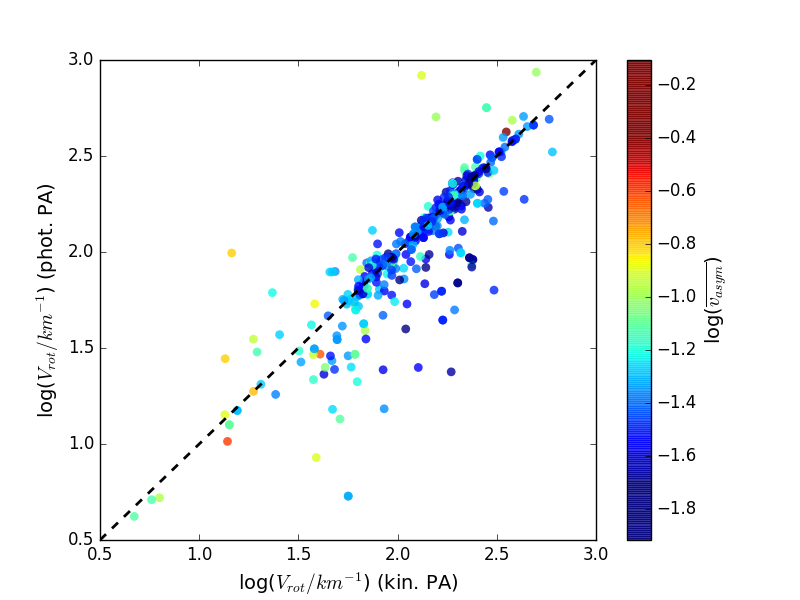}
\caption{$V_{rot}$ using the photometric against kinematic slit, with points coloured by asymmetry. Using the photometric slit produces a systematic underestimation of $V_{rot}$, as compared with the kinematic slit.}
\label{fig:vsvsvs}
\end{figure}

\begin{center}
\begin{table}
  \begin{tabular}{ b{1.5cm}  l  b{1.5cm}  l b{1cm l }}
    \hline
    Classification & Fraction of misaligned galaxies & Number  \\ \hline
    Face-on & 9$\%$ & 10 \\
    Barred & 29$\%$ & 30\\
    Asymmetric & 24$\%$ & 25\\
    Normal & 37$\%$ & 39\\
    \hline
  \end{tabular}
\caption[Table caption text]{This table gives results of a by-eye classification of misaligned galaxies in order to identify causes of misalignment. Face-on galaxies may have uncertain photometric PAs, whereas bars can lead to spurious photometry. Asymmetric galaxies may have variable or distorted PAs and kinematics. Lastly, `normal' galaxies are both photometrically and kinematically normal, but are nevertheless misaligned.} 
\label{table:paclass}
\end{table}
\end{center}

\begin{figure}
\includegraphics[width=9cm]{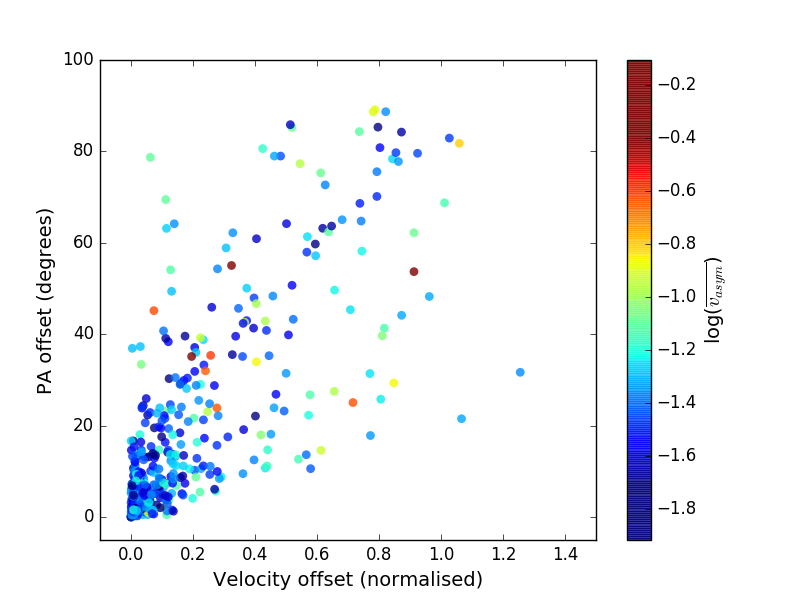}
\caption{Difference between $V_{rot, kin.PA}$ and $V_{rot, phot.PA}$, {normalised by $V_{rot, kin.PA}$}, against offset in kinematic and photometric PA, with points coloured by asymmetry. There is a correlation between difference in calculated $V_{rot}$ and PA offset, and between asymmetry and PA offset. }
\label{fig:vel_pa_offset}
\end{figure}

Fig.~\ref{fig:vel_pa_offset} demonstrates this relationship (with a correlation $\rho=0.66, p=1\times10^{-15}$) between difference in $V_{rot, kin.PA}$ and $V_{rot, phot.PA}$ and PA offset (with $V_{rot, kin.PA}$ and $V_{rot, phot.PA}$ being the rotation velocities calculated using the kinematic and photometric PAs, respectively). We define velocity offset as:
\begin{equation}
\mathrm{V_{offset}}={\mathrm{V_{rot, kin.PA}}-\mathrm{V_{rot, phot.PA}}}
\end{equation} 

There is a further correlation between PA offset and asymmetry, with $\rho=0.17, p=1\times10^{-4}$. As shown in Fig.~\ref{fig:bias}, there is also a relationship in the 2D spatially resolved TFR between scatter off the TFR and PA offset.

Fig.~\ref{fig:p_a_s} shows the three key relationships in this work: $\overline{v_{asym}}$ against PA offset, $\overline{v_{asym}}$ against residuals above and below the TFR, and PA offset against scatter off the 2D spatially resolved TFR for all galaxies in  Fig.~\ref{fig:twod_tfr_asyms}, with Table~\ref{table:p_a_s} giving partial Spearman rank correlation test results. Note that galaxies scattered above the TFR with large PA offsets are asymmetric and barred galaxies. Although the inverse relationship between $\overline{v_{asym}}$ and scatter off the TFR is the strongest, the correlation between $\overline{v_{asym}}$ and PA offset and the inverse relationship between scatter off the TFR and PA offset are also evident. The relative strengths of these relationships point to kinematic asymmetry as the strongest driver of scatter off the TFR.

However, the influence of PA choice on scatter is significant. Our photometric PA slit method is similar to placing a long slit at the photometric PA to calculate rotation velocity, as is seen in literature [e.g. \citet{neistein1999tully,pizagno2007tully,reyes2011calibrated,simons2015transition}]. Some scatter in the TFRs in these works can be attributed to their use of the photometric PA. Deeper photometry, or targeting more nearby galaxies, may be able to reduce scatter in these cases, as the derived PA would be less influenced by bars, and galaxies identified as belonging to some categories in Table~\ref{table:paclass} could be excluded.

\begin{center}
\begin{table}
  \begin{tabular}{ b{4cm}  l  b{1.8cm}  l b{1.8cm} l }
    \hline
     & $\rho$ & p-value  \\ \hline
    $\overline{v_{asym}}$/ PA offset (! TFR scat.)& $ 0.19$ & $1.2\times10^{-7}$\\
    $\overline{v_{asym}}$/ TFR scat. (! PA offset)& $-0.27$ & $1.9\times10^{-14}$\\
    PA offset/TFR scat. (! $\overline{v_{asym}}$) & $-0.12$ & $2.1\times10^{-2}$ \\
    \hline
  \end{tabular}
\caption[Table caption text]{This table gives partial correlation parameters between $\overline{v_{asym}}$, PA offset and scatter off the TFR, with the variable held constant indicated by `!'.  Tests were performed using all points in Fig.~\ref{fig:twod_tfr_asyms}. Whilst the relationships persist in all three cases, the links between PA offset and $\overline{v_{asym}}$ and scatter off the TFR and $\overline{v_{asym}}$ are stronger than that between PA offset and scatter off the TFR. } 
\label{table:p_a_s}
\end{table}
\end{center}

\section{Conclusions}
\label{sec:conclusion}

Without selecting based on asymmetry or Hubble type, we show the TFR for 729 kinematically and morphologically diverse galaxies in the SAMI Galaxy Survey sample and find:
\begin{equation}
\mathrm{log}(V_{rot}/kms^{-1})=0.31 \pm 0.092 \times \mathrm{log}(M_{*}/M_{\odot})- 0.93 \pm 0.10
\end{equation}

Scatter increases at low stellar mass, but the TFR, particularly for low asymmetry galaxies, is still present. Scatter below the TFR is well correlated with asymmetry across the stellar mass range of the sample. Asymmetry is inversely correlated with stellar mass, confirming the result of \citet{bloom2016sami}.

We find that slit-based, rather than 2D spatially resolved, measurements of rotation velocity increase scatter below the TFR. Further, if the slit is placed at the photometric PA, rather than the kinemetric, scatter increases again. Scatter is thus highly dependent on PA selection. This result may be relevant to other TFR works, particularly those aiming for the tightest possible TFR for use as a distance indicator.



\section*{Acknowledgements}

The SAMI Galaxy Survey is based on observations made at the Anglo-Australian Telescope. The Sydney-AAO Multi-object Integral field spectrograph (SAMI) was developed jointly by the University of Sydney and the Australian Astronomical Observatory. The SAMI input catalogue is based on data taken from the Sloan Digital Sky Survey, the GAMA Survey and the VST ATLAS Survey. The SAMI Galaxy Survey is funded by the Australian Research Council Centre of Excellence for All-sky Astrophysics (CAASTRO), through project number CE110001020, and other participating institutions. The SAMI Galaxy Survey website is http://sami-survey.org/.

GAMA is a joint European-Australasian project based around a spectroscopic campaign using the Anglo-Australian Telescope. The GAMA input catalogue is based on data taken from the Sloan Digital Sky Survey and the UKIRT Infrared Deep Sky Survey. Complementary imaging of the GAMA regions is being obtained by a number of independent survey programs including GALEX MIS, VST KiDS, VISTA VIKING, WISE, Herschel- ATLAS, GMRT    ASKAP providing UV to radio coverage. GAMA is funded by the STFC (UK), the ARC (Australia), the AAO, and the participating institutions. The GAMA website is http://www.gama-survey.org/.

Support for AMM is provided by NASA through Hubble Fellowship grant $\#$HST-HF2-51377 awarded by the Space Telescope Science Institute, which is operated by the Association of Universities for Research in Astronomy, Inc., for NASA, under contract NAS5-26555.

SMC acknowledges the support of an Australian Research Council Future Fellowship (FT100100457).

M.S.O. acknowledges the funding support from the Australian Research Council through a Future Fellowship (FT140100255).

{We thank the referee for their helpful and insightful comments, which clarified and improved the paper.}


\bibliography{paper_III}
\bibliographystyle{mn2e}
\end{document}